\documentclass[namedreferences]{solarphysics}

\usepackage[hyperref,optionalrh]{spr-sola-addons} 
\usepackage{graphicx}        
\usepackage{amssymb}         
\usepackage{amsmath}         
\usepackage{color}           
\usepackage{breakurl}       
\usepackage{hyperref}
\hypersetup{
    colorlinks=true,
    linkcolor=blue,
    filecolor=blue,      
    urlcolor=blue,
}
\usepackage{amssymb}      

\graphicspath{ {/} }
\usepackage{verbatim}
\usepackage{float}
\usepackage{enumitem}
\usepackage{mathtools}

\begin{document}
\begin{article}
\begin{opening}

\title{Infrared Imaging Spectroscopy Using Massively Multiplexed Slit-Based Techniques and Sub-Field Motion Correction}

%
\author[corref,addressref=aff1,email ={schad@nso.edu}]{\inits{T.A.}\fnm{Tom}~\lnm{Schad}}  
\author[addressref=aff2,email ={lin@ifa.hawaii.edu}]{\inits{H.}\fnm{Haosheng}~\lnm{Lin}}  
\address[id=aff1]{National Solar Observatory$^{*}$, 8 Kiopa`a Street, Pukalani, HI 96768, USA}
\address[id=aff2]{Institute for Astronomy, University of Hawai`i, Pukalani, Hawaii 96768, USA}
\runningauthor{Schad \& Lin}
\runningtitle{Seeing-compensated multiplexed spectroscopy}


\begin{abstract}
Targeting dynamic spatially-extended phenomena in the upper solar atmosphere, a new instrument concept has been developed and tested at the Dunn Solar Telescope in New Mexico, USA, that provides wide-field, rapid-scanning, high-resolution imaging spectroscopy of the neutral helium $\lambda10830$ spectral triplet. The instrument combines a narrowband imaging channel with a novel, co-spatial, 17 parallel-long-slit, grating-based spectrograph that are simultaneously imaged on a single HgCdTe detector. Over a $170'' \times 120''$ field of view, a temporal cadence of 8.5 seconds is achieved between successive maps that critically sample the diffraction limit of the Dunn Solar Telescope at 1083 nm ($1.22 \lambda/D = 0.36''$) and provide a resolving power ($R = \lambda / \delta\lambda$) up to $\sim40000$ with a 1 nm bandwidth (\textit{i.e.}, 275 km sec$^{-1}$ Doppler coverage). Capitalizing on the strict simultaneity of the narrowband channel relative to each spectral image (acquired at a rate of 10 Hz), this work demonstrates that sub-field image motion introduced by atmospheric seeing may be compensated in each mapped spectral data cube.  This instrument furnishes essential infrared spectral imaging capabilities for current investigations while pioneering techniques for high-resolution, wide-field, time-domain solar astronomy. 
\end{abstract}


\end{opening}


\section{Introduction} \label{sec:intro}

Solar astronomy relies on multi-dimensional spectroscopic techniques \citep[see, \textit{e.g.},][]{bershady2009} to remote sense the dynamic, interconnected, thermal and magnetic scales of the solar atmosphere.  In comparison to many night-time applications, science objectives require greater resolution and near-simultaneous coverage of spatial, spectral, and temporal domains \citep[see, \textit{e.g.,}][]{socas_navarro2010}.  The immense solar intensity permits relatively denser observational sampling, but challenges remain, including optimizing sampling and domain coverage, resolving solar evolutionary time scales, achieving high image quality in the presence of atmospheric seeing, and reaching the extreme signal-to-noise demands of polarimetric measurements \citep{lagg2015}.  Additionally, the stratified nature of the solar atmosphere necessitates multi-wavelength observations of multi-thermal diagnostics.  These disparate demands and technological limitations naturally spur trades in instrument design.\footnote{For a recent review of current and planned solar instrumentation, see \cite{kleint2015}.} 

Currently limited is support for high-resolution imaging spectroscopy of important near-infrared (NIR) spectral diagnostics, \textit{e.g.}, He I $\lambda$10830 and Fe I $\lambda$15648,  despite scientific advantages \citep[see, \textit{e.g.},][]{penn2014}.  Instead the majority of solar imaging spectrographs are Fabry-P\'{e}rot, \textit{i.e.}, etalon, based, fast-tuning spectral imagers operating at visible wavelengths \citep{von_der_luhe2000, cavallini2006, reardon2008, scharmer2008}.  While some disadvantages exist, a strength of etalon based systems is that rapidly acquired image stacks can be used to boost image quality via post-facto corrective techniques such as destretching,\footnote{Destretching refers to a range of methods that enforce smooth variation of the spatial content image relative to a reference image by doing sub-field local correlation and interpolation.} speckle interferometry \citep{woeger2008}, and blind deconvolution \citep{van_noort2005}.  Unfortunately, the development of NIR etalon-based spectrographs has trailed their visible counterparts, in part due to the later availability of fast readout large-format NIR detectors and the adaptation of commercial etalons \citep{cao2004}.  The Near-InfraRed Imaging Spectropolarimeter (NIRIS) instrument \citep{cao2006,cao2012} at the Big Bear Solar Observatory is the first to extend Fabry-P\'{e}rot technology to NIR wavelengths and has demonstrated observations of the Fe I $\lambda15648$ photospheric spectral line.  No other comparable NIR imaging spectrographs are currently known. 

Motivated by the study of chromospheric and cool coronal dynamics, the current work pursues slit-based alternatives for NIR imaging spectroscopy, in particular to target the He I $\lambda$10830 triplet, which is an important spectral line formed in the upper solar atmosphere and is especially useful as a polarized diagnostic of chromospheric magnetic fields \citep[see, \textit{e.g.,}][]{harvey1971, schad2016, leenaarts2016} .  While long-slit grating-based spectrographs excel at high precision spectral measurements, temporal resolution and image quality typically suffer due to the time required to map an image and the role of atmospheric seeing.  However, two techniques, when used in tandem, may alleviate these limitations.  First, spatially multiplexed grating-based spectrographs, which greatly reduce image mapping time, are fairly simple to implement with the use of parallel long-slits and narrowband filtering.  In solar astronomy, multi-slit methods were pioneered by \cite{martin1974} and further developed by \cite{srivastava1999} and \cite{jaeggli2010}.  More recently, \cite{lin2014} introduced the highly efficient Massively Multiplexed Spectroheliograph (mxSPEC) concept targeting synoptic full-disk solar applications. Second, methods capable of improving the image quality of slit-based instruments do exist, though not often used.  \cite{johanneson1992} minimized seeing-induced distortions of single-slit scanned maps by applying, to the spectra, destretch parameters calculated from co-registered high rate context images.  More powerful speckle deconvolution methods for slit-scanned spectrograms were later introduced by \cite{keller1995} and show significant promise. Spatially multiplexed spectroscopy, when combined with post-facto image correction, may empower slit-based imaging spectroscopy to achieve performance similar to Fabry-P\'{e}rot (FP) approaches while yielding some advantages, as discussed below.

Here, we advance the massively multiplexed technique of \cite{lin2014} for the application of high-resolution He I $\lambda10830$ imaging spectroscopy of the chromosphere and cool corona.  We evolve \citeauthor{lin2014}'s mxSPEC design to achieve higher spatial, spectral, and temporal resolution, while using an advanced adaptive optics system and a post-facto correction similar to \cite{johanneson1992} to boost image quality of the mapped spectrograms. Below we describe the instrument setup, calibration methods, the seeing distortion correction, and results for scientific targets and then discuss the advantages of these techniques.

\begin{figure*}  
\centering
\includegraphics[width=0.975\textwidth,clip=]{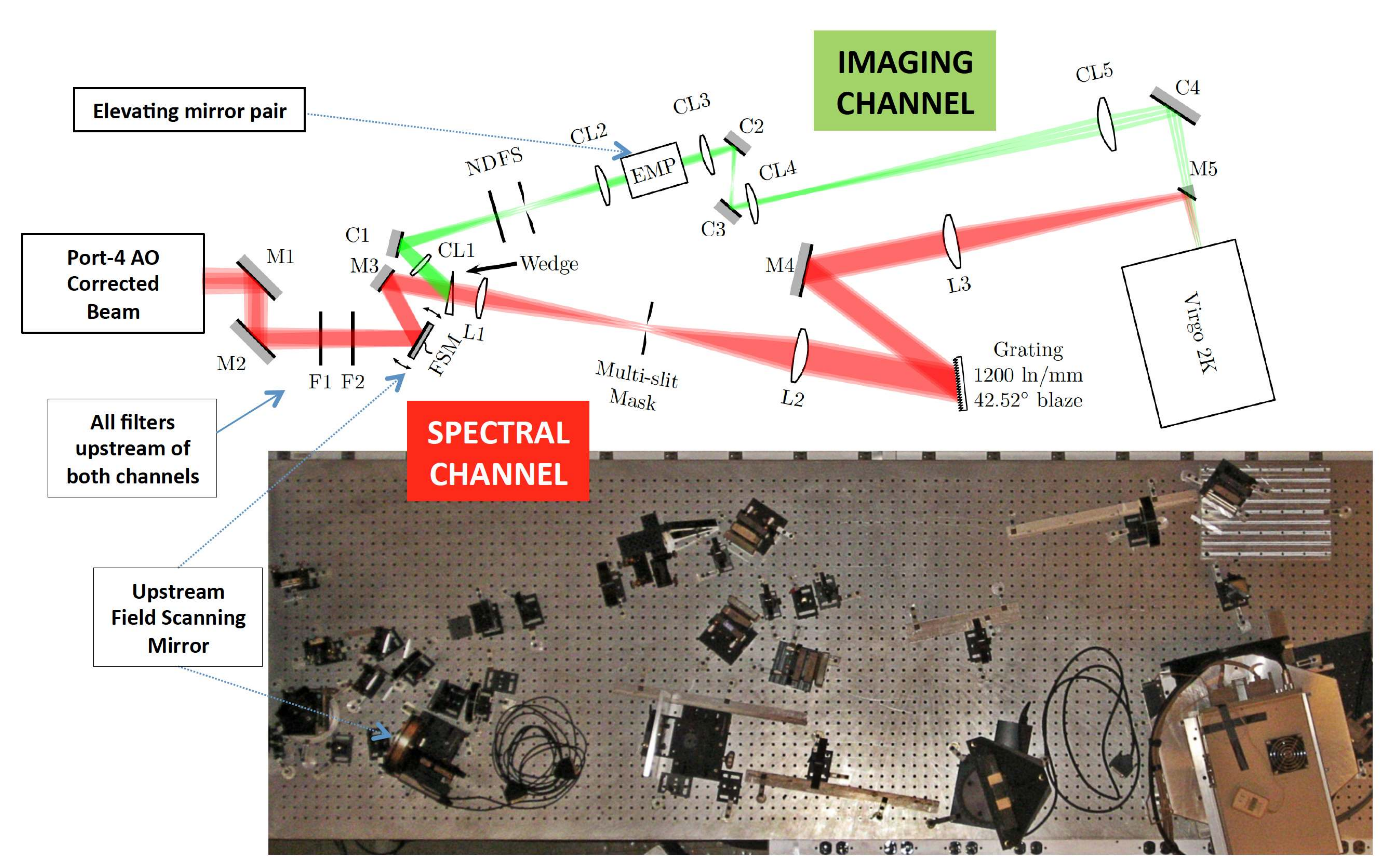}
\caption{Optical layout (\textit{top}) and top-view photo (\textit{bottom}) of the dual-arm imaging multi-spectrograph.  The optical beam enters from the left as it exits the upstream DST port No. 4 high order adaptive optics (HOAO) system and spectral beamsplitters.  The red optical path denotes the primary beam that enters into the multi-slit spectral channel. The green optical path shows the narrowband imaging channel that is split from the primary beam via a wedge.  The relative component distances are shown to scale.  See text for details.}
\label{fig:optics}
\end{figure*}

\section{Methods}\label{sec:methods}

\subsection{Optical layout}\label{sec:optics}

The multi-slit imaging spectrograph has been designed to fit within the envelope of a $91 \times 305$ cm optical bench permanently installed on port No. 4 of the rotating platform inside the 76 cm aperture Dunn Solar Telescope \citep[DST:][]{dunn1964,dunn1991}.  Figure~\ref{fig:optics} shows the optical layout and a top-view photo of the installed instrument without its coverings.  Ahead of the instrument are the High Order Adaptive Optics (HOAO) system \citep{rimmele2004} and a dichroic beam splitter that reflects visible light to the Interferometric BiDimensional Spectrometer \citep[IBIS:][]{cavallini2006,reardon2008}, which is a FP based visible imaging spectrograph.  The horizontally-oriented beam provided to the multi-slit spectrograph is collimated with an angular magnification of 35 (\textit{i.e.,} $\varnothing21.66$ mm exit pupil).

As this is a prototype instrument, it is built from existing optomechanical components available at the DST.  The design goals were to maximize field coverage (limited by the telescope port's $195''$ square field stop), critically sample the diffraction limit at He I $\lambda10830$ ($1.22 \lambda/D = 0.36''$), and provide a minimum spectral resolving power ($R = \lambda / \delta\lambda$) of $\sim30000$, consistent with sampling schemes used by previous high-resolution studies of chromospheric/coronal dynamics \citep[see, \textit{e.g.},][]{vissers2012}.  The spectral bandpass is constrained by the availability of a high transmission 1.3 nm -25 dB bandwidth bandpass isolation filter centered at the Si I $\lambda10827$ spectral line, which, while not optimal, provides adequate coverage of the He I $\lambda10830$ triplet. 

In order to implement post-facto seeing correction similar to \cite{johanneson1992}, the instrument is comprised of two channels: a narrowband imaging channel and a multi-slit spectral channel, both of which are directed towards and imaged on separate halves of a 2048 x 2048 Raytheon Virgo HgCdTe focal plane array (pixel pitch is 20 $\mu$m).  This ensures the spectral and narrowband images are precisely cotemporal.  The detector is housed in a liquid nitrogen chilled dewar build by the University of Hawai$`$i.  As shown in the schematic (Fig.~\ref{fig:optics}), collimated light exiting the upstream dichroic beam splitter is transferred by two flat mirrors (M1 and M2) and directed through an order-sorting filter (F1) and the narrow 1.3 nm bandpass isolation filter (F2) before an intermediary telescope pupil is focused onto an actuator controlled field steering mirror (FSM). After a fourth reflection (M3), the beam is split by an Oriel anti-reflection coated wedge, wherein approximately 1\% of the light is directed towards the imaging channel (green in figure), and the rest towards the grating-based multi-slit spectrograph.  
	
\begin{table}
\caption{Instrument properties of the seeing-compensated multi-slit spectrograph}
\label{tbl:instrument}
\begin{tabular}{ll}
\hline 
\bf{Field scanning properties}         &              \\
Common field of view                   & $185'' \times 130''$ \\
Step size                              & $0.187''$  \\
Number of steps                        & 65                  \\ 
Average mapping cadence                & $\sim$ 8.5 seconds  \\ 
\hline
\bf{Spectrograph Channel}              &              \\
Spectral lines                         & Si I $\lambda$10827; He I $\lambda$10830 \\
Number of illuminated slits            & 17           \\
Projected slit width                   & $0.187''$  \\ 
Projected slit length                  & $155''$  \\
Sampling along slit                    & $0.153''$  \\ 
Anamorphic magnification          & 1.56         \\
Spectral pixel width ($\Delta\lambda$) & 120.117 m$\AA$             \\ 
Theoretical resolving power            & $\sim41000$  \\
Slit free spectral separation          & 1.34 nm      \\
Filtered spectral bandwidth            & 1.08 nm        \\
Doppler coverage                       & -195 to +85 km sec$^{-1}$             \\
\hline
\bf{Imaging Channel}                   &              \\
Spatial sampling                       & $0.128''$     \\ 
\hline
\end{tabular}
\end{table}

Collimated light entering the spectrograph channel is focused by a 400 mm focal length (FL)  doublet onto a mask with 49 parallel long slits.  Each slit has a width of 12.5 $\mu$m and is separated by 750 $\mu$m, as in \cite{lin2014}.  The measured image scale at the slit is $14.7''$ mm$^{-1}$; thus, each slit samples $0.187''$ of the solar image.  Post-slit is a classical refractive spectrograph consisting of a 400 mm FL collimating doublet (L2), a 165 mm wide 1200 lines mm$^{-1}$ (42.52$^{\circ}$ blaze) reflective diffraction grating, and an 800 mm FL doublet (L3) acting as the camera lens. A fold mirror (M4) is placed between the grating and L3, while a small triangular mirror (M5) is placed between L3 and the detector.  The spectrograph is operated in first order, with an incident angle of 28.3$^{\circ}$.  Slit-induced diffraction ensures a sufficient illumination width ($\sim72$mm) to provide resolving power above the limit imposed by sampling and the convolution of the slit image.  The 2:1 reimaging optics and anamorphic magnification (1.56) result in an exit slit separation of 2.34 mm ($\sim117$ pixels) and a spectral dispersion of 120.117 m\mbox{\AA} pixel$^{-1}$, thus achieving a maximum bandwidth between slits of 1.4 nm.  The final calculated resolving power is 41000.  17 illuminated slits, with a total angular separation on sky of $180''$, fit on the detector.  Spatial sampling along the slits is 0.15$''$ pixel$^{-1}$, meaning $153''$ can be placed within the bottom half of the Virgo 2K detector.

\begin{figure}    
\centerline{\includegraphics[width=0.85\textwidth,clip=]{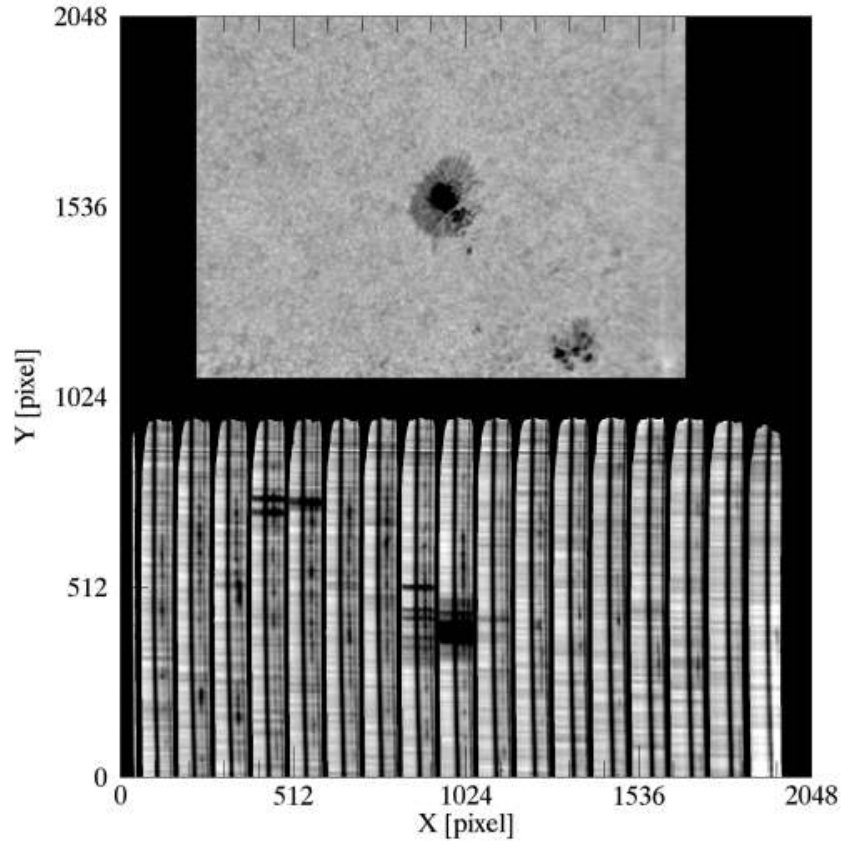}}
\caption{Dark and flat field corrected focal plane image of the multi-slit instrument showing the context image (\textit{top}) and the 17 illuminated slit spectra (\textit{bottom}).  The solar target is NOAA active region 12465, observed on 9 December 2016.  Note that the spectral image, which is rotated 180$^{\circ}$ with respect to the constant image, remains fixed during the image scan while the context image shifts in the horizontal (X) direction.}
\label{fig:focal_image}
\end{figure}

The imaging channel, operating on the wedge's collimated reflected light, is first focused by a 400 mm FL doublet (CL1). A field stop is placed at the focus to occult a laterally displaced ghost image created by the wedge.  Ahead of the focus is a fold mirror (C1) and an 0.15 optical density neutral density filter used to prevent detector saturation.  To place the context image on the top half of the detector, the height of the optical path must be raised by half of the FPA physical height.  To accomplish this without large beam conflicts, we demagnify the pupil image using a 200 mm FL doublet, which results in a 10.833 mm pupil image.  Two closely spaced fold mirrors are then used in an elevating configuration (EMP: `elevating mirror pair') to achieve the 20.48 mm beam lift.  Following the EMP, a 1:1 relay optical system consisting of twin 120 mm FL doublets (CL3 and CL4) and two fold mirrors (C2 and C3) is used to deviate the beam direction and transfer the pupil telecentrically.  Finally, the beam passes through a 490 mm FL doublet, is folded by a mirror (C4), passes over the small M5 mirror, and comes to a focus on the top half of the FPA.  M5 vignettes approximately 15 percent of the bottom portion of the context image.

An example focal plane image with dark and flat field calibrations applied is displayed in Figure~\ref{fig:focal_image}.  The top portion of the image contains the narrowband context image while the bottom half captures the 17 slit spectral images. The context image edges have been removed as flat-fielding errors are introduced by time variable translational image stroke at wide field angles caused by the deformable mirror of the HOAO system.  Within the slit spectral images, the most prominent spectral feature is the dark Si I $\lambda$10827 photospheric line with the He I triplet and a H$_{2}$0 telluric line at $\lambda10832$ immediately redward (to the right in image). A summary of the instrument properties is given in Table~\ref{tbl:instrument}.

\subsection{System control and operation}\label{sec:operation}

To map the solar surface, the FSM scans the solar image across the slits in coordination with image acquisition.  For control, script-based software written in the Interactive Data Language (IDL$^{\tiny\textregistered}$)\footnote{IDL$^{\textregistered}$ is a product of Exelis Visual Information Solutions, Inc., a subsidiary of Harris Corporation (Exelis VIS).} interfaces to C libraries communicating with the detector control electronics and a Newport ESP 300 driving the FSM linear actuators.  During one scan, the detector records 65 exposures at its fastest frame rate (9.53 Hz, \textit{i.e.,} 105 ms integration time), with one exposure used for each FSM step, which has an angular size equal to the projected slit width.  The FSM is commanded to move at the beginning of each new frame, and its move time is a small fraction of total integration time.  Five extra steps ensure full field coverage and some overlap for adjacent slits.  An advantage of placing the FSM ahead of both the spectral and imaging channels is that the image scanning step size is self-calibrated by the translated narrowband image on the detector.  It also allows better removal of bad pixels in the context image stack via interpolation. At the end of each scan, the FSM returns to its starting position while the acquired data block (520 MB) is swapped to a new memory address for FITS\footnote{FITS stands for Flexible Image Transport System, an open standard for storing scientific images.} file creation and archiving via a parallel IDL process.  The total overhead for control and data writing is $\sim 1.7$ seconds per scan; thus, repeated scans are acquired at a cadence of 8.5 seconds. 

\subsection{Calibration and data reduction}\label{sec:data_analysis}

\subsubsection{Detector response}

Dark, bias, and non-linear characteristics of the detector response are calibrated using established methods.  Dark and bias are removed by the subtraction of the average of a set of unilluminated frames acquired with the same exposure time. Unfortunately, the bias for the Virgo 2K exhibits some temporal fluctuations that manifest differently on its 16 output channels.  This is a limitation imposed by the lack of correlated double sampling for this detector.  For low illumination applications, \textit{e.g.} off-limb observations, the fluctuations can be corrected by enforcing continuity of the bias level across different rows of the detector, but on-disk applications suffer from this artifact.  We derive the non-linear response of the detector by measuring its photon response curve under various illumination conditions.  This is done on a column by column basis as the readout mode of the detector leads to a variable gain during the column by column frame read. The detector also has a great number of randomly distributed bad pixels that are interpolated using nearest-neighbor averaging for correction.

\subsubsection{Flat-fielding}

For the removal of pixel to pixel electronic and optical gain fluctuations, two sets of flat field observations are acquired:  solar flatfields and lamp flatfields.  Solar flatfields consist of a set of 50 full field scans (\textit{i.e.} the same operation described in Section~\ref{sec:operation}) acquired at a nominal position near solar disk center and away from any solar activity.  During acquisition the telescope performs a smooth, continuously scanning, small amplitude random guide procedure that repoints the telescope away from the nominal flat field target position.  This aids in the averaging of solar structures present in the solar flat field scans.  Lamp flatfields are acquired by injecting a facility provided Quartz Tungsten Halogen lamp near the port No. 4 prime focus and with a matched focal ratio.  Lamp flat fields do not contain the spectral lines that exist in the solar flatfield and thus facilitate their removal.

To construct the applied flatfield, an average solar flatfield field scan is first determined from the 50 acquired scans after the detector response has been corrected.  For the narrowband imaging portion of the scan, this average serves as the applied flat field.  For the spectral portion, we must remove the spectral lines prior to applying the flat to the data. The spectral lines are isolated from the narrowband filter profile by dividing by the average lamp flatfield; however, due to small deviations in the alignment of the lamp relative to the solar beam, the bandpass in the lamp flat is slightly shifted relative to the solar flat field.  Before division, we use cross-correlation along each spectral row in each slit spectral image to derive a shift that is then applied to the lamp flat field profile to best match the profile shape in the solar flatfield. Then we extract the spectra from each individual slit and correct for the in-plane spectrograph-induced spectral line curvature by shifting each spectral row to align the positions of the spectral lines.  The amplitude of the slit curvature is approximately 11 spectral pixels.  The average spectrum along each slit is then used to remove the spectral lines from each spectrum of that slit in the original average solar flatfield scan. To do this, we reapply the slit curvature to the average spectrum for each spectral row and slit and apply it to the original solar flatfield scan directly by division.  When the line-less flat field is applied to science data, the result is as shown in Figure~\ref{fig:focal_image}. 

\subsubsection{Spectral dispersion}

The average flat-field spectrum along each slit (discussed above) is also used to determine the linear spectral dispersion and wavelength scale by measuring the line center positions of the Si I 10827.089 $\mbox{\AA}$ and telluric H$_{2}$O line at 10832.108 $\mbox{\AA}$ as in \cite{kuckein2012}.  No corrections for orbital motions are applied.  The measured spectral dispersion is on-average 120.117 m\mbox{\AA} pixel$^{-1}$; though, there is a $\pm 2.5\%$ variation of the dispersion of the outer slits compared with the central slit on account of the variation of the effective littrow angle of each slit.  In addition, the spectral bandpass shifts by $\sim 4.2$ \mbox{\AA} across the $185''$ field of view due the pupil placement of the narrowband filter.  While we do not measure the achieved spectral resolution, it is less than the theoretical resolving power for the system on account of limited focusing precision during alignment and unknown optical performance of available optics.  While the data is scientifically important, any detailed spectral line profile measurements using this demonstration data must take into account the non-benign line spread function, which can be determined, \textit{e.g.}, by comparision of the flat field spectra with a high-quality spectral atlas. 

\subsubsection{Image-spectra angular coregistration}\label{sec:registration}

\begin{figure}    
\centering
\includegraphics[width=0.65\columnwidth,trim = 0.75cm 0.1cm 0.5cm 0.05cm, clip ]{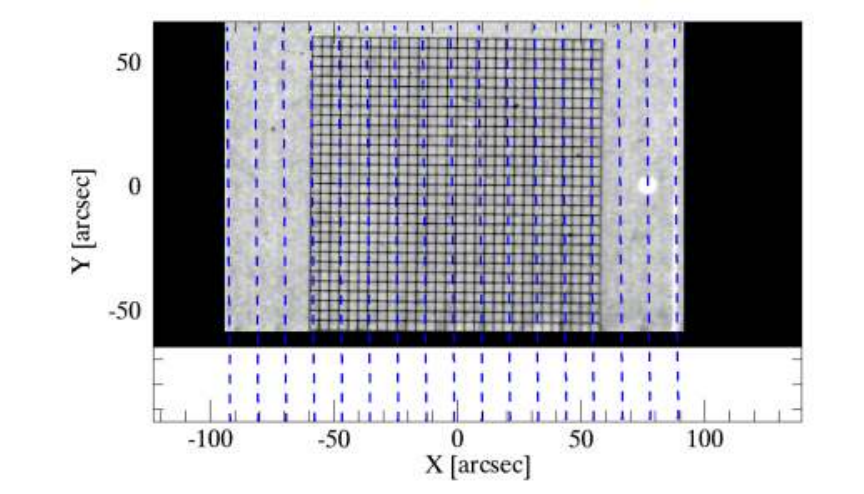}
\caption{Location of coregistered long slits overplotted on fixed frequency line grid target at its central field scanning position.  The slits sample more of the field than the narrowband context image.  Angular coordinates are referenced to the telescope boresight.}
\label{fig:slit_registration}
\end{figure}

Angular registration of the slit-scanned spectral image and the narrowband image is facilitated by observations of a fixed frequency line grid target placed at the telescope's prime focus (see Figure~\ref{fig:slit_registration}).  First, the angular scale in each channel is determined by measuring the average line spacing in the acquired line grid observations and using the known plate scale of the telescope's prime focus (3.76$''$ mm$^{-1}$).  For the spectral channel, only the image scale along the slit is determined, and we assume the line grid is orientated orthogonal to the slits.  The measured narrowband channel image scale ($p_{c}$) is found to be 0.128$''$ pixel$^{-1}$, while the spatial sampling along the slit in the spectral image ($p_{\lambda}$) is 0.153$''$ pixel$^{-1}$.  

Next we coregister the two channels using scanned observations of the line grid target.  Due to the fixed position of the slits relative to the detector, there exists a geometric mapping between pixels within the spectral images and corresponding detector pixels within the narrowband imaging channel.  However, since the FSM's scanning direction is not strictly perpendicular to the slits and exhibits small variations in scan step width, the relationship between the scan-reconstructed image and the narrowband image cannot directly be written via a polynomial or affine transformation.  Rather we must correct for the FSM inhomogeneities prior to deriving the transformation. 

The frames acquired by the imaging channel during an individual scan permit a calibration of the FSM motion.  We represent the raw imaging channel coordinates as $\{u_{c},v_{c}\}$, where `c' stands for context, and write the set of $n$ imaging frames as $f_{c}(u_{c},v_{c})|_{n}$, with $n=0,1,\dots,64$.  Similarly, we can represent the scan-reconstructed spectral image of the line grid (at pseudo-monochromatic wavelength $\lambda$) as $f_{\lambda}(u_{\lambda},v_{\lambda})|_{slit}$, where $u_{\lambda}$ is the step number coordinate (\textit{i.e.}, $0,1,\dots,64$), $v_{\lambda}$ is the vertical axis pixel index, and $slit$ denotes which slit is being registered. To correct for FSM motion, scalar translational alignment of the $f_{c}|_{n}$ images relative to the center scan position image ($f_{c}|_{32}$) are determined via cross-correlation.  The average step size along the $u_{c}$ axis is $\bar\Delta_{step} = 1.46$ pixels $(0.187'')$ while the \textit{average} motion along the $v_{c}$ axis is assummed to be zero.  Deviations from the average scanning motion as a function of n are given by $\{\delta u_{c},\delta v_{c}\}|_{n}$. 

FSM motion inhomogeneities are corrected in the scan-reconstructed spectral images by remapping the coordinates of $f_{\lambda}$ using the FSM scanning deviations $\{\delta u_{c},\delta v_{c}\}|_{n}$.  The transformed image $f_{\lambda}'(u_{\lambda}',v_{\lambda}')$ uses mapping functions given by $u_{\lambda}' = u_{\lambda} - \delta u_{c}|_{u_{\lambda}} / \bar\Delta_{step}$ and $v_{\lambda}' = v_{\lambda} + \delta v_{c}|_{u_{\lambda}} \cdot (p_{c}/p_{\lambda})$.  Note $u_{\lambda}$ and $n$ are equivalent variables and that the sign of the second term relates to the relative parity of the two instrument channels.

A transformation is next derived between the FSM motion corrected scan-reconstructed spectral image $f_{\lambda}'$ and the center scan position imaging channel image ($f_{c}|_{32}$) via cross-correlation and first order polynomial mapping functions (\textit{i.e.}, rotation, scaling, and translation).  Writing the transformed spectral image as $g_{\lambda}'(x_{c},y_{c})$, the mapping functions are given by $x_{c} = \sum_{i=0}^{N} \sum_{j=0}^{N} P_{ij}u_{\lambda}'v_{lambda}'$ and $y_{c} = \sum_{i=0}^{N} \sum_{j=0}^{N} Q_{ij}u_{\lambda}'v_{lambda}'$.  We use the \textit{auto\_align\_images.pro} routine available in the \textit{Solar Software} \citep[SSW:][]{freeland1998} IDL distribution to optimize this transformation.

The two transformations discussed above provide a mapping between the slit image coordinates $u_{\lambda},v_{\lambda}$ and the imaging channel coordinates $u_{c},v_{c}$ for each spectrograph slit.  Using these mappings, the location of the slits relative to the narrowband image acquired at the center FSM position is shown in Figure~\ref{fig:slit_registration}.  Note that the spectrograph slits cover approx 30$''$ more field in the Y direction than the imaging channel.  In the subsequent analysis below, we ignore this portion of the field.  

\section{Seeing compensation}\label{sec:seeing_comp}

\begin{figure}    
\centering
\includegraphics[width=0.45\columnwidth,clip=]{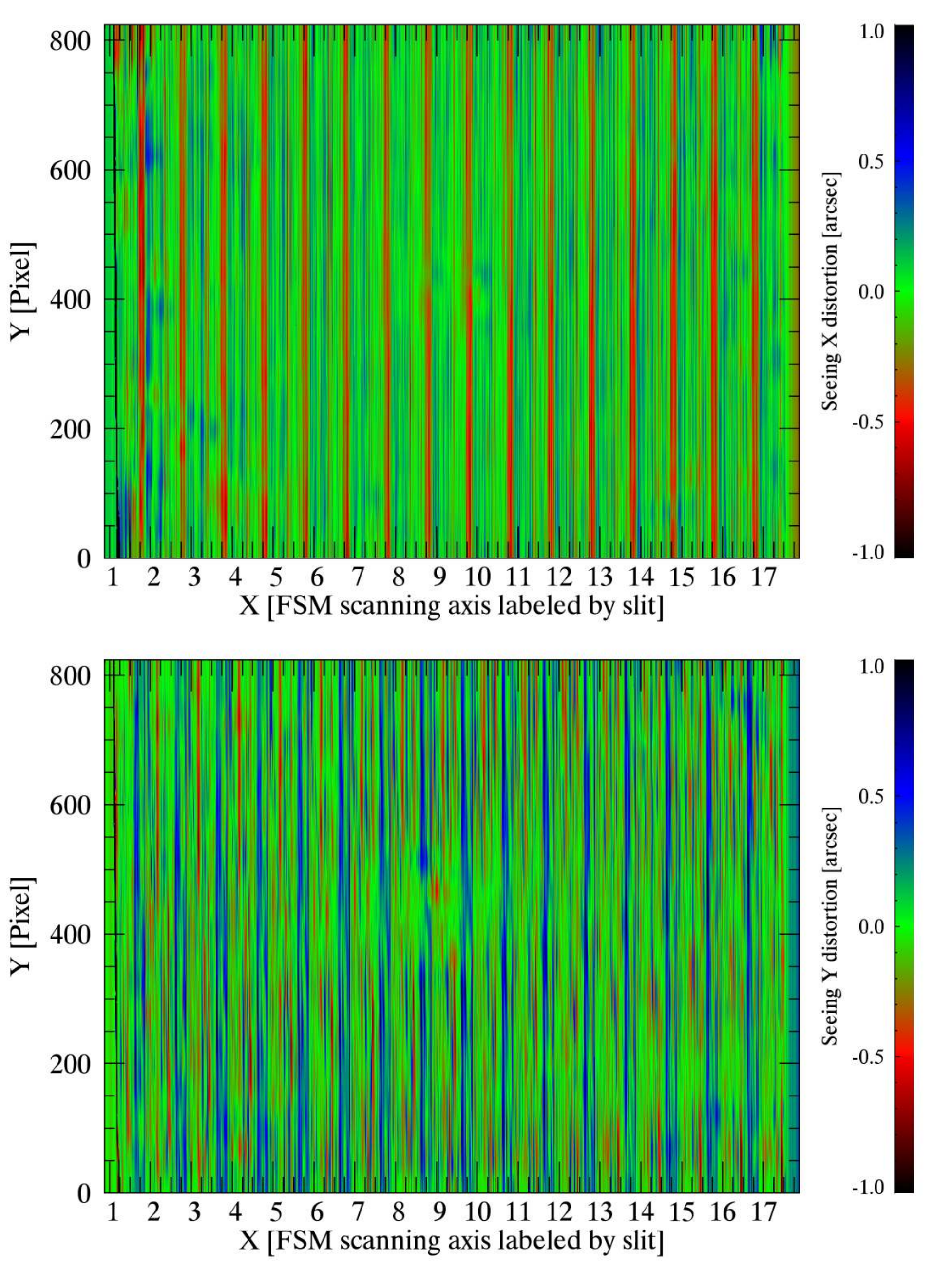}  
\includegraphics[width=0.45\columnwidth,clip=]{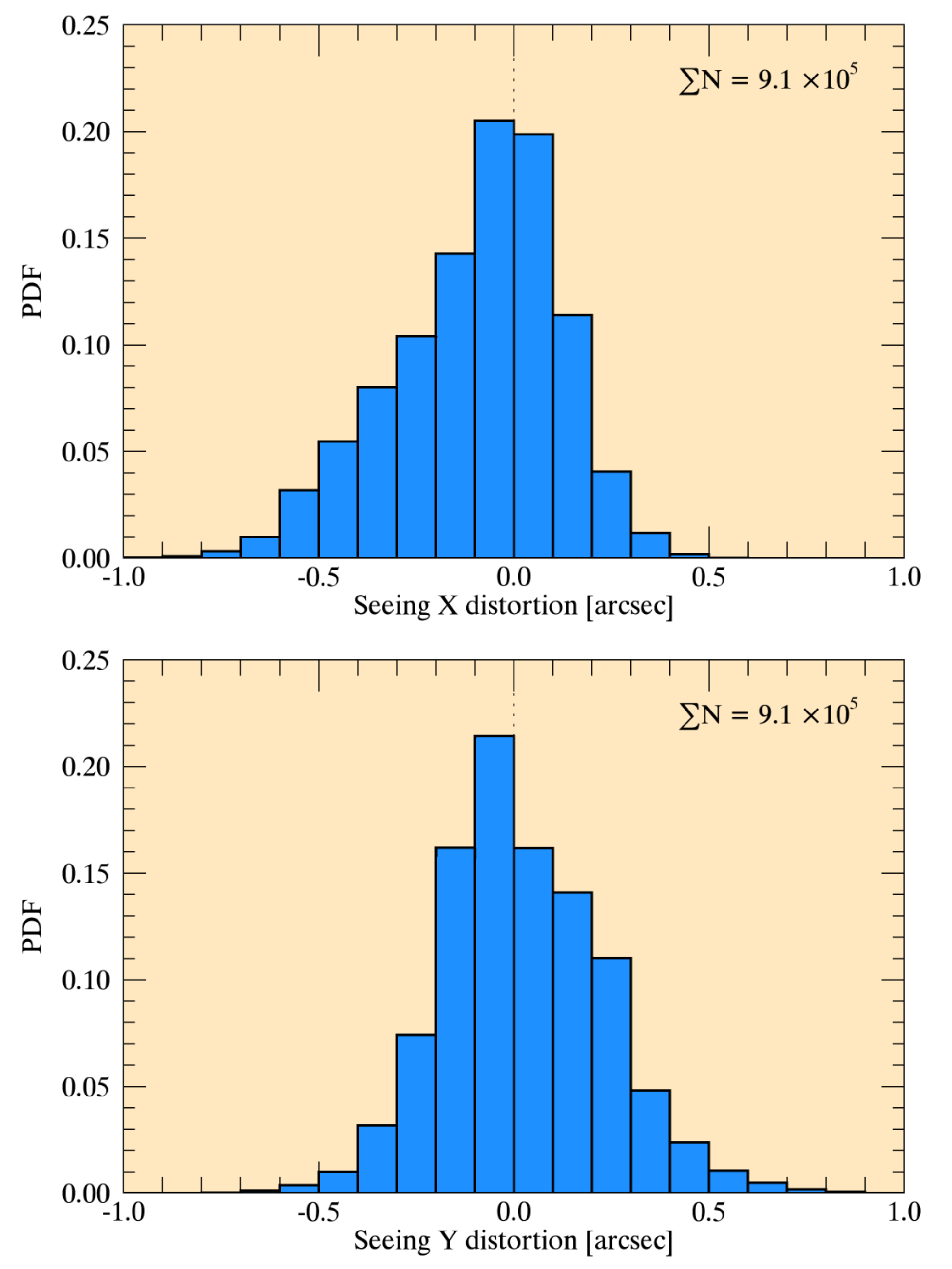}  
\caption{(\textit{left}) Maps of the derived seeing distortion in each direction for one scanned map (during of 8.5 seconds) and as sampled by the 17 long slits.  The y axis refers to pixels along the slit. The x axis corresponds to the FSM stepping axis ($65$ steps $\times 17 = 1105$) but each of the 65 steps are labelled only by the associated slit number. (\textit{right}) Histograms of the angular seeing distortions across the field of view.}
\label{fig:seeing_deltas}
\end{figure}

Careful registration between the spectral and imaging channels facilitates compensation of differential sub-field image motion introduced by atmospheric seeing in each spectral data cube.  This zeroth order distortion is actively mitigated by the HOAO system; however, the HOAO system operates at visible wavelengths and corrects well only over a small isoplanatic patch ($<10''$).  Post-facto destretching algorithms \citep{november1988,rimmele1994} are often used to compensate residual displacements within imaging data sets, and have been applied to single-slit spectral data by \cite{johanneson1992}.  These algorithms compute local sub-field correlations of subsequent images relative to a reference frame and then interpolate each frame to a uniform spatial grid.  In order to \textit{freeze} the seeing distortion during the exposure, integration times need to be short compared to the seeing correlation time ($\tau_{0} \cong r_{0}/v$, where $r_{0}$ is the Fried parameter and $v$ is wind speed of the atmospheric turbulence layer, see, \textit{e.g.}, \cite{rimmele2011}). $r_{0}$ scales as $\lambda^{6/5}$, so there is an advantage to observing in the infrared.  For typical observing conditions ($r_{0} @ 500 nm = 10$ cm; $v = 10$ m/s), the time constant at 1083 nm is $\sim25$ msec, roughly a factor of 4 shorter than our integration time of 105 ms, which is the shortest exposure length supported by the Virgo 2K detector.  

We employ the destretch algorithms developed by \cite{rimmele1994} to determine the residual image displacements in the narrowband imaging channel image-set, \textit{i.e.}, $f_{c}(u_{c},v_{c})|_{n}$, and then apply these to the spectral data cube using the transformation parameters derived above.  The imaging coordinates are first mapped to angular coordinates on the sky, ignoring sub-field motion.  These coordinates, \textit{i.e.}, $\{x_{sky}',y_{sky}'\}$, are related to $\{u_{c},v_{c}\}$ for each image, n, via the plate scale, the average FSM scanning step, the FSM deviations $\{\delta u_{c}, \delta v_{c} \}|_{n} $, and two additional parameters $\{ \Delta x_{c}, \Delta y_{c}\}|_{n}$ corresponding to variations in the telescope tip/tilt or pointing control, which for on-disk targets under HOAO control are typically benign.  After aligning the image-set $f_{c}|n$ with the primed sky coordinates, we produce a reference image from the running average of 30 consecutive images (3.15 seconds in duration).  The destretching process then provides mapping arrays (see Figure~\ref{fig:seeing_deltas}) between the reference image, whose coordinates $\{x_{sky},y_{sky}\}$ are taken to represent the true solar scene, and the interim sky coordinates $\{x_{sky}',y_{sky}'\}$.  The destretching procedure uses an iterative approach with progressing smaller sub-field sizes ($12''$, $8''$,and $4''$). \footnote{The same correction has been applied to off-limb targets; however, the correction typically can only support larger sub-field kernel sizes dependent upon available off-limb structure in the narrowband image.} Following this step, we attain a complete set of transformation functions between the spectral data cube coordinates and the destretched angular coordinates. 

Using the derived seeing-compensated image transformation mapping functions, we remap the calibrated spectral data cube into the corrected sky coordinates with the same angular scale as the context image (\textit{i.e.}, $0.128''$ pixel$^{-1}$).  As sub-field image motion results in an irregularly gridded set of coordinates; we first triangulate the transformed data coordinates using a Delaunay triangulation \citep{lee1980}, and then linearly interpolate onto a regular grid using the triangulated coordinates.\footnote{The IDL$^{\tiny\textregistered}$ procedure \textsc{TRIGRID} is used for the interpolation.}  This procedure can also identify voids in the corrected data coordinates in the event that residual image displacements are significantly larger than the image scale, \textit{e.g.}, during periods of bad seeing or poor HOAO performance. 

\section{Results}\label{sec:results}

\begin{figure}    
\centering
\includegraphics[width=0.975\textwidth,clip=]{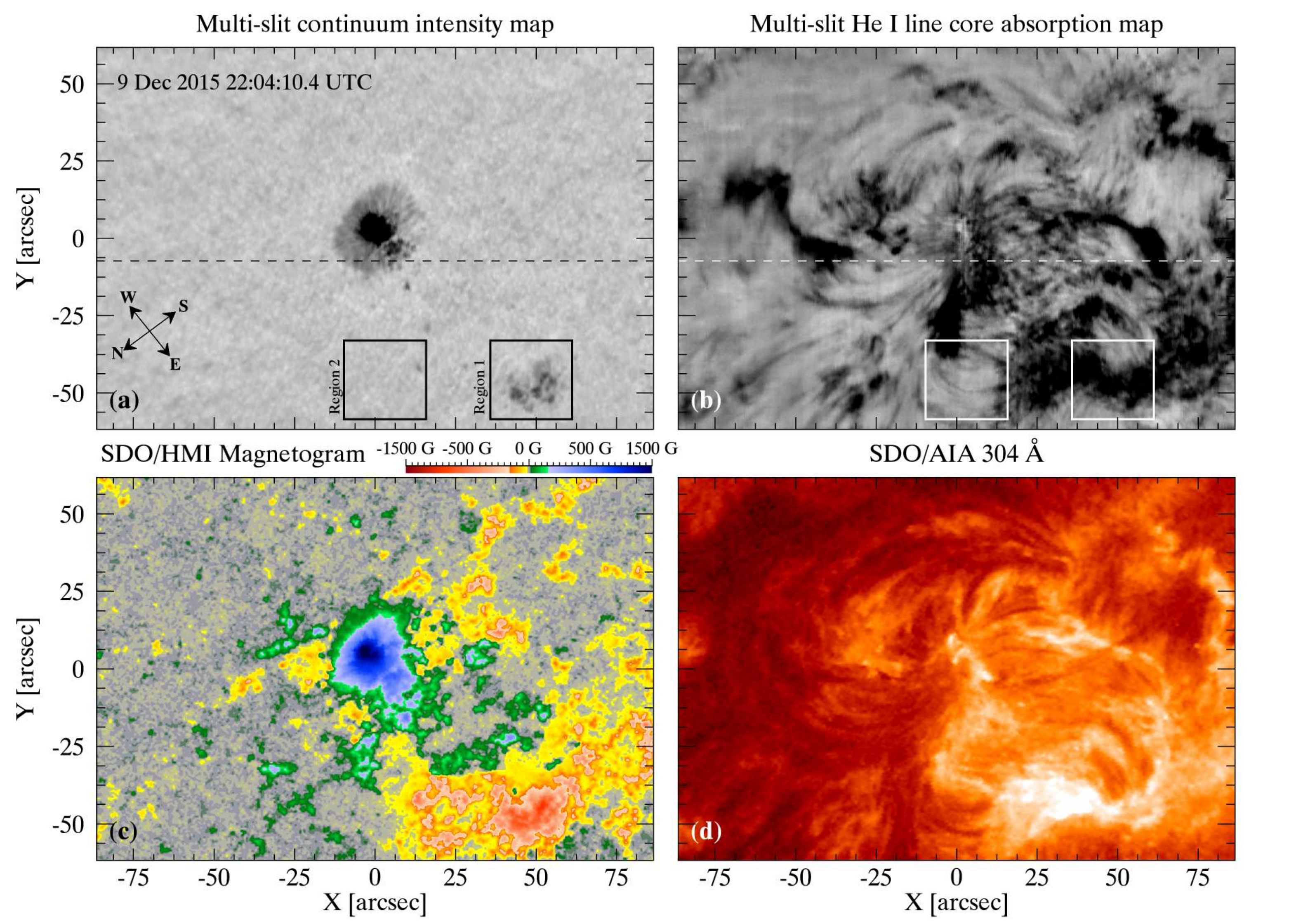} \\
\includegraphics[width=0.975\textwidth,clip=]{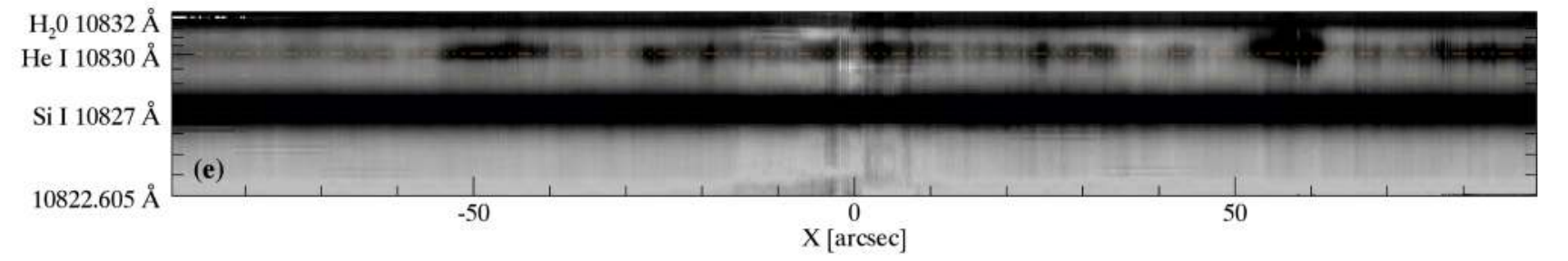}
\caption{Multi-slit imaging spectroscopy observations acquired 9 Dec 2015.  Panels (a) and (b) show maps of the continuum intensity and the normalized He I line core intensity resulting from a single spectral image data cube (duration of 8.5 seconds) extracted from a 55 minute time series.  The imaging quality of each has been improved by the seeing compensation technique described in the text.  Co-aligned SDO/HMI magnetograms and 304 $\AA$ images from SDO/AIA are shown in panels (c) and (d).  He \textsc{I} spectra extract along the dashed line in panels (a) and (b) is shown in panel (e).  Two sub-regions identified in the spectral maps are further analyzed in Figures~\ref{fig:region1} and~\ref{fig:region2}. \textit{An animation of panels a-d is available in the ESM.}}
\label{fig:on_disk}
\end{figure}

Figure~\ref{fig:on_disk} and its associated animation available in the electronic supplementary material (ESM) display observations acquired by the multi-slit imaging spectrograph between 21:25 and 22:20 UTC on 9 December 2015 that demonstrate the instrument's unique capabilities for wide-field imaging spectroscopy of the He I $\lambda10830$ triplet.  The target is NOAA Active Region 12465 in the southern hemisphere and consists of a large leading sunspot and a smaller trailing multi-umbrae region. Solar north is directed toward the bottom left of the field as shown.The figure and its animation also show co-aligned photospheric magnetograms from SDO/HMI \citep{scherrer2012} and 304 $\AA$ EUV images from SDO/AIA \citep{lemen2011}.  The dynamics of NOAA AR 12465 in the chromosphere and lower corona are clearly evident in the animation, and, in addition, a portion of a filament eruption originating from the nearby NOAA AR 12464 is visible in both 304 $\AA$ and He I $\lambda10830$ (see upper right of field between 21:45 and 22:20 UTC).

\begin{figure}    
\centering
\includegraphics[width=0.975\textwidth,clip=]{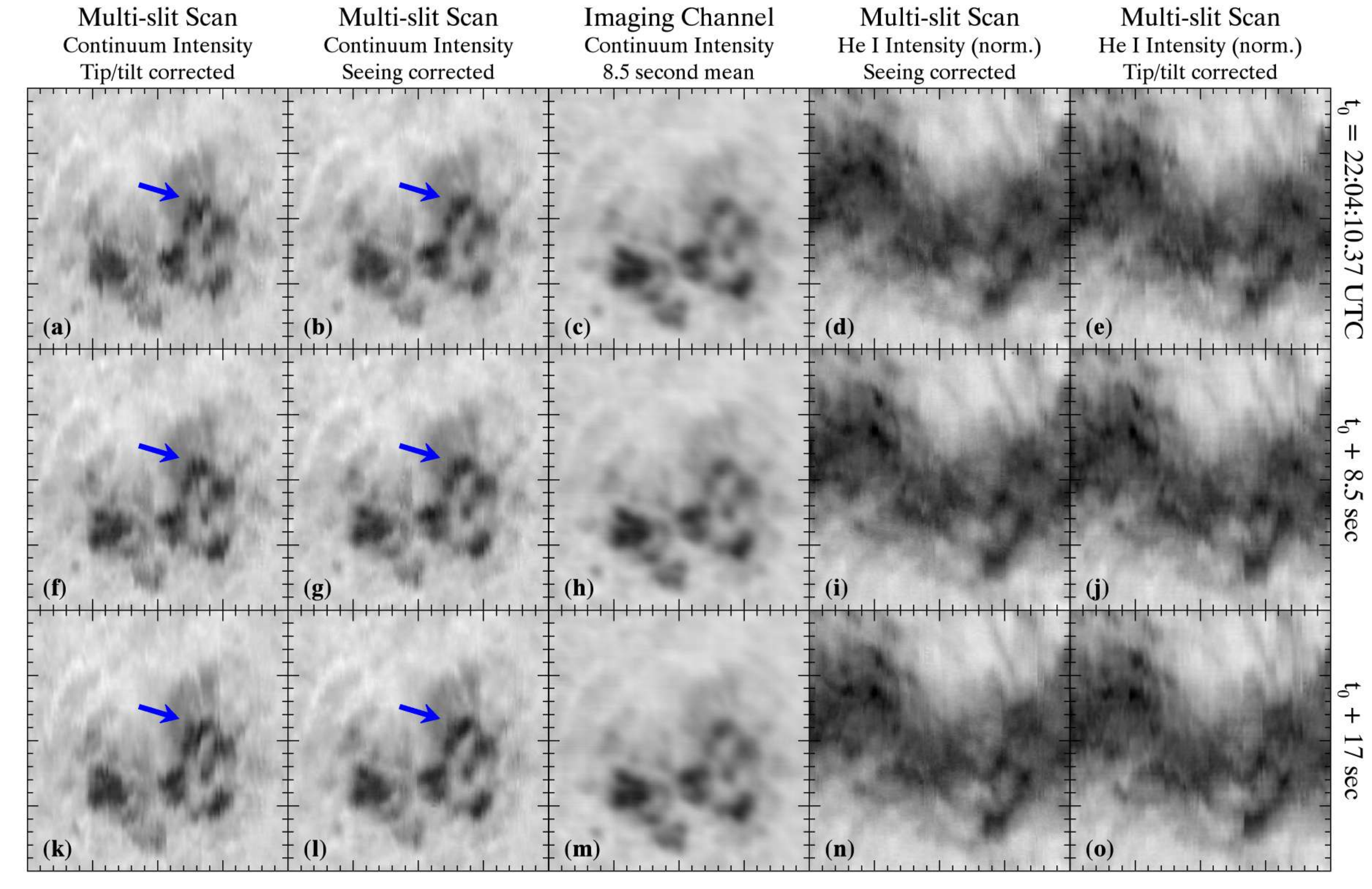} \\
\caption{Comparison of the uncorrected and seeing-compensated data for Region 1 indentified in Figure~\ref{fig:on_disk}.   From left to right, the plots show maps of the continuum intensity gleaned from the spectral data and corrected only for tip/tilt, the same maps corrected for tip/tilt and sub-field seeing motion, a mean context image acquired by the imaging channel of the instrument during the scan, the sub-field seeing corrected spectral maps of normalized He I line core intensity, and the same maps corrected only for tip/tilt.  \textit{An animation is available in the ESM.}}
\label{fig:region1}
\end{figure}

\begin{figure}    
\centering
\includegraphics[width=0.975\textwidth,clip=]{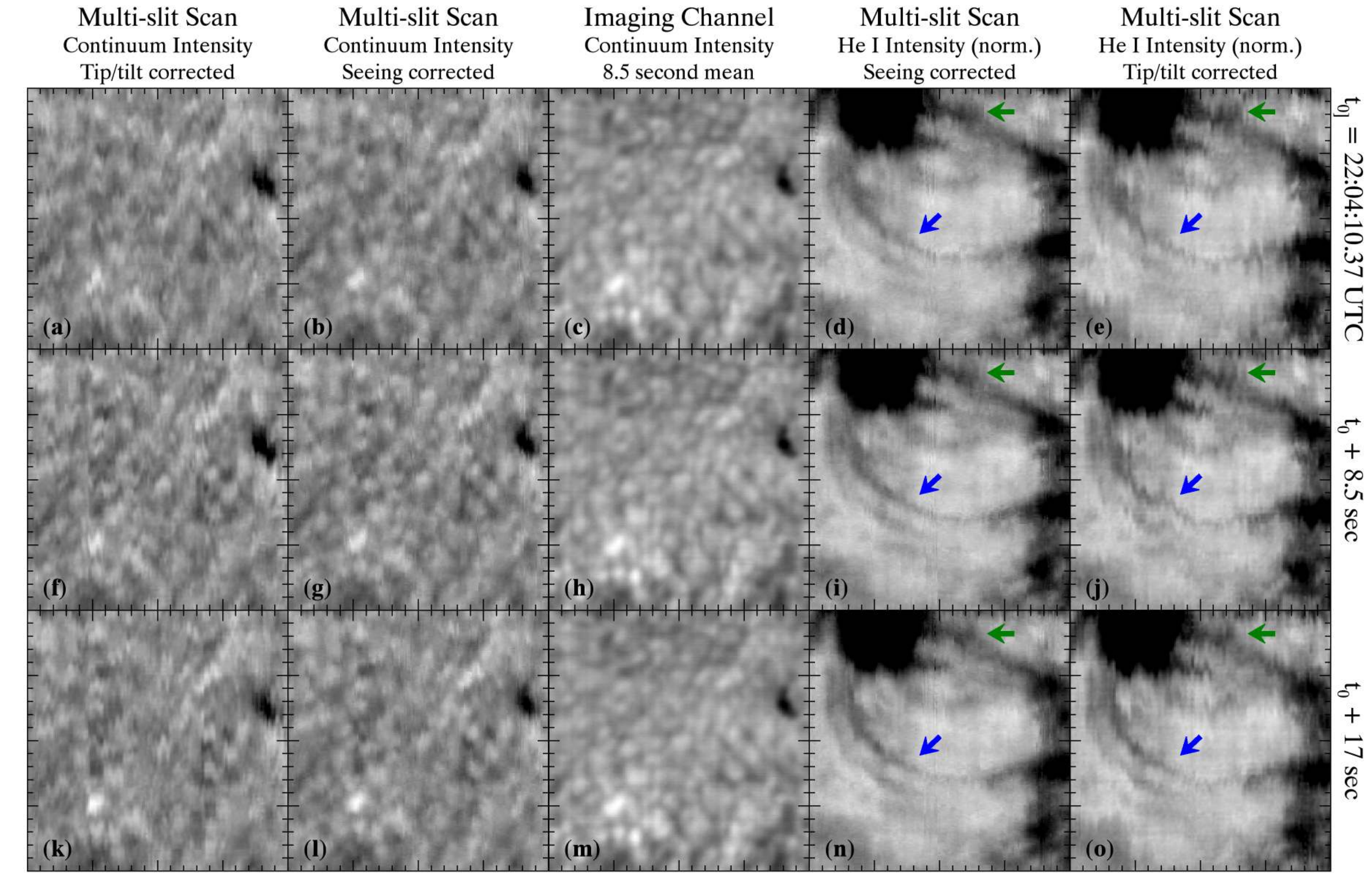} \\
\caption{Same as in Figure~\ref{fig:region1} but for Region 2.  \textit{An animation is available in the ESM.}}
\label{fig:region2}
\end{figure}

The seeing-compensated continuum intensity and He I absorption maps derived from the spectral scans (Figure~\ref{fig:on_disk} panels a and b) extend across the majority of the active region, which is an advantage of the relatively large field of view of this instrument, and exhibit improved image quality compared to the uncompensated slit-scan maps, which are compared in detail below for the two sub-regions in Figures~\ref{fig:region1} and~\ref{fig:region2}.  In the continuum map, the evolutionary granulation pattern is evident in the quiet sun surrounding the sunspot in the continuum map.  Although the map is not of diffraction-limited quality due to higher order seeing degradation, spatial dislocations introduced by seeing motion have been significantly minimized.

Example spectra extracted from the data cube along the dashed line in Figure~\ref{fig:on_disk} panels (a) and (b) are normalized by their continuum intensity and displayed in panel (e).  These spectra are derived from 17 different long slits, each with its own slit curvature and small variation in dispersion; yet, the spectra exhibit continuity across the field of view.  The bright feature in the middle of the image is introduced by the detector bias fluctuations discussed above, the negative impacts of which can also be seen in the animated spatial maps.  The apparent He I velocity structure, and in particular large red-shifts at $X = 55''$, result from the strong dynamics of the upper atmosphere and signify the importance of He I imaging spectroscopy.  Preliminary sequences of derived Dopplergrams, which are currently under investigation, feature large filament flows as well as quickly evolving fibril/spicule dynamics. 

Observing conditions on 9 December 2015 were variable with high level cirrus clouds and often strongly turbulent atmospheric seeing. The example data shown here consist of 373 field scans performed during the $\sim55$ minute time series.  Sub-field motion is significantly minimized by the compensation technique during periods of moderate seeing.  However, when conditions degrade and the contrast of features in the context image decline, the seeing compensation technique does not perform well.

To demonstrate the improvements gained from the seeing compenstation technique, we compare the corrected and uncorrected maps of the two sub-regions identified in Figure~\ref{fig:on_disk}. Region 1 contains the multi-umbrae region of the regions's trailing polarity and is located approximately $70''$ away from the center of the field of view where the limited field of view adaptive optics system achieves its lock-point.  For three consecutive field scans each separated by 8.5 seconds, Figure~\ref{fig:region1} shows maps of the continuum intensity gleaned from the spectral data and corrected only for tip/tilt\footnote{Tip/tilt correction refers to the application of the values $\{ \Delta x_{c}, \Delta y_{c}\}|_{n}$ as defined in Section~\ref{sec:seeing_comp}.  Seeing-correction (or sub-field coorection) refers to the mapping between $\{x_{sky}',y_{sky}'\}$ and $\{x_{sky},y_{sky}\}$ provided by the destretching procedure and applied to the spectral data.} (\textit{i.e.}, panels a,f,k), the same maps corrected for tip/tilt and sub-field seeing motion (b,g,l), a mean context image acquired by the imaging channel of the instrument during the scan (c,h,m), the sub-field seeing corrected spectral maps of normalized He I line core intensity (d,i,n), and the same maps corrected only for tip/tilt.  An animation is also available within the ESM.  Away from the AO lock-point, the target field experiences large seeing-induced differential motion.  In the tip/tilt corrected maps, this sub-field motion results primarily in a false jagged character of the scanned features (see region indicated by blue arrow).  The corresponding seeing-compensated maps shows a significant improvement in spatial integrity as confirmed by the favorable comparison with the context narrowband image.  Note that the context image in this region is slightly out of focus compared with the spectral map.  As time progresses, the seeing variations introduce different distortions to the region (see a,f,k), which are in each case compensated by our technique (b,g,l). 

Likewise, spatial dislocations are significantly improved for the He I absorption maps, which is best illustrated by inspection of the chromospheric structure apparent in Region 2 as shown in detail within Figure~\ref{fig:region2} (and its assocated animation in the ESM). Low-lying chromospheric loops show smooth topologies in the seeing corrected maps  (panels d,i,n) in contrast to the jagged structure apparent in the maps corrected only for tip/tilt (e,j,o).  See the features indicated by the arrows in the figure.  Note that sub-region labeled "Region 2" does not contain significant photospheric structure other than granulation.  Thus, the seeing-compensation technique primarily depends on quantifying the seeing-induced motion of the photospheric granulation for correction, and indeed the jagged structures in the tip/tilt corrected maps of continuum intensity (a,f,k) are considerably reduced in the seeing-corrected maps (b,g,l).  As low-lying chromospheric loops are associated with a presumably aligned magnetic field, they are generally expected to be smooth curvilinear structures.  Magnetohydrodynamic (MHD) waves may induce transverse motions; however, for the loops shown in Figure~\ref{fig:region2}, it is clear that the atmospheric seeing has induced the jagged structure and not MHD waves.  This emphasizes the importance of the seeing-compenstion technique to scientific interpretation in the chromosphere. 

\section{Discussion}\label{sec:discussion}

The results of this work testify to the potential for using seeing-compensated massively multiplexed slit spectroscopy to achieve high-resolution observations of the solar atmosphere.  In particular this instrument provides capabilities necessary for the study of dynamics over large regions in the upper atmosphere using the important He I triplet.  Here we assess the instrument capabilities and compare these with Fabry-Perot (FP) based instruments.  We also discuss prospects for the future development of image quality improved massively multiplexed spectrographs.

Performance limitations aside, given a non-evolving multidimensional data cube and an equal number of detector pixels, all imaging spectroscopy techniques are equivalent if they make efficient use of available detector pixels.   With the introduction of time-variance in the data and technological limitations, various techniques have different advantages and disadvantages meaning cost and performance must be weighed against particular science objectives.  In our case, the need for simultaneous high temporal resolution and large spectral coverage to capture dynamic phenomena, such as erupting filaments and coronal rain, finds advantages in multiplexed slit spectroscopy.  As compared to etalon-based (\textit{i.e.}, Fabry-Perot (FP) based) instruments, the number of exposures per scan is not set by the number of spectral samples.  Our instrument obtains 100 spectral samples over its field-of-view via a 65 exposure sequence.  A conventional single-channel solar FP instrument \citep[see review by][]{kleint2015} requires a minimum of 100 exposures to achieve the same spectral coverage.  Assuming comparable throughput, our instrument can obtain a temporal cadence roughly a factor of 2 faster for the same signal-to-noise. 

In addition to Doppler velocity coverage surrounding highly-resolved spectral lines, solar astronomy requires spectral diversity, that is, the observation of multiple spectral regions/lines.  Our current measurements achieve spectral diversity via co-temporal visible light imaging spectroscopic observations provided by IBIS.  However, for scientific cases requiring high-temporal cadence, \textit{e.g.,} temporally resolved observations of coronal rain \citep{schad2016agu}, we must limit the number of visible spectral lines observed by IBIS in order to achieve the requisite temporal resolution.  Single-channel FP instruments can only observe different spectral lines in sequence, which limits spectral diversity for time-domain physics.  While our multi-slit instrument only observes the spectral window near He I 1083 nm, grating-based spectrographs, including multiplexed designs, can more easily support simultaneous multi-spectral observations using a single dispersing optic \citep[see., \textit{e.g.},][]{jaeggli2010,lin2014,lin2016}.  

This work has demonstrated that the spatial integrity of multiplexed imaging spectral data can be significantly improved via post-facto seeing-compensation, yet the achievable image quality of etalon-based instruments remains superior.  However, a significant challenge for FP design in the era of large $>4$ meter class solar telescopes is the required FP aperture.  As discussed by  \cite{beckers1998}, \cite{von_der_luhe2000}, and \cite{kentischer2012}, the pupil apodization effect for FPs in a telecentric mounting configuration, which from the point of view of spatial resolution is preferred over the collimated mounting, introduces velocity errors that scale with the target contrast and the beam focal ratio incident of the FPs.  Typical solar requirements for velocity errors (\textit{10s of m/sec}) and field-of-view ($\sim >50''$) consequently require very large FPs, \textit{i.e.} clear apertures greater than 250 mm for a 4 meter telescope, which are extremely challenging to manufacture  \citep{sigwarth2016}. 

Existing technologies and design heritage likely make multiplexed diffraction-grating-based spectrographs more appealing and cost-effective for wide-field imaging spectrographs at large-aperture solar facilities if image quality can be improved.  This is largely due to conventional commerically-available echelle gratings, with sizes typically as large at 300 x 400 mm, being able to provide sufficient resolving power for wide-field high-resolution spectroscopy at significantly reduced cost compared with etalons.  Even so, spectrograph design remains challenging and throughput optimization and the need to ease optimal design for large-aperture systems with typically fast focal ratios and tight space envelopes spurs the need for larger gratings.   Such optics have been previously constructed by mosaicing together conventional gratings \citep[see., \textit{e.g.},][]{pilachowski1995, dekker2000}.  

Both etalon and slit-based instruments are routinely used to conduct full Stokes polarized spectroscopy, which is an essential tool in remote sensing solar magnetic fields.   The large light collecting area of large-aperture solar telescopes is in part motivated by the need for high polarimetric sensitivity and accuracy.  Grating spectrographs must be carefully designed and calibrated in order to avoid errors caused by the well-known grating induced polarization discussed by \cite{breckinridge1971}.  Meanwhile, for large focal ratios, \cite{doerr2008} found that air gapped FPs have a fairly benign influence on polarization; though, polarization artifacts arise when FPs are used in faster optical beams.  Typically, however, it is time-dependent systematics that limit polarization sensitivity, in particular polarized interference fringes \citep[see][for review]{semel2003}.  Post-facto methods for fringe removal are more routinely used by slit-based instruments, as in \cite{casini2012}; however, as the spectral bandwidth is narrowed (as is the case for FP instruments), such methods become more challenging to implement. 

Etalon and slit-based instruments both require scanning, or tuning, to build up a single observation of a multidimensional data cube.  For science targets that evolve spatially and spectrally on shorter timescales than the scanning period, both instruments are limited.  Only integral field unit (IFU) spectrographs, which simultaneously observe the spectral and two-dimensional spatial domains, overcome this limit.  In solar physics, IFUs are challenging to build that achieve high spatial and spectral resolution over a significant field of view.  Though, much progress has been made, and solar IFUs are currently being developed for a number of solar projects \citep{lin2006,calcines2014,schad2014,katsukawa2016, suematsu2016}.

The advantage of the massively multiplexed spectrograph is its simple cost-efficient design.  Yet, key to the success of slit-based imaging spectrographs for use in high spatial resolution investigations are methods to improve image quality as we have advanced here.  Our results show that the spatial integrity of the data can be significantly improved post-facto.  These results might be further improved by speckle reconstructing the context image quality instead of using a temporal average; however, to reach diffraction-limited quality, more advanced methods need to be pursued.  The method of \cite{keller1995} offers one post-facto option, while multi-conjugate adaptive optics  \citep{schmidt2017} may soon yield improved wide-field active correction that reduces the need for post-facto correction.  In either case, high-performance large-format focal plane arrays with high speed readout are needed that can obtain the data within a time period shorter than the characteristic time scale of atmospheric seeing.


\section{Summary}\label{sec:conclusion}

A seeing-compensated multi-slit imaging spectrograph has been advanced as a post-focus instrument and tested at the Dunn Solar Telescope in New Mexico, USA.  This instrument furnishes the ability to observe chromospheric dynamics in the He I triplet at $\lambda$10830 at high resolution over a large field of view.  Its use of simultaneous on-band context imaging permits post-facto improvement of the spatial image quality of slit-scanned spectral images.  \cite{schad2016agu} recently used this instrument to investigate He I spectral intensity distribution of coronal rain, which is a phenomena that is difficult to target in a statistically meaningful manner without rapid large field spectroscopy that this instrument provides.

Such alternative approaches to seeing-compensated imaging spectroscopy, \textit{i.e.}, ones that do not require large tuning etalons or integral field units, may provide cost-effective solutions to multi-spectral observations at next-generation large-aperture telescopes, in particular at infrared wavelengths.  Currently, the National Science Foundation's 4-meter aperture Daniel K. Inouye Solar Telescope \citep{rimmele2015}, is under construction in Maui, Hawai`i.  It will provide dynamic imaging spectroscopy at visible wavelengths with a cutting edge etalon based instrument \citep{schmidt2016} and at infrared wavelengths via an $80 \times 60$ field-point integral field unit \citep{lin2006,schad2014}.  Multi-slit spectroscopy, when paired with post-facto image quality improvements, offer an alternative means to achieve expanded high-resolution field coverage for multiple spectral lines simultaneously at rapid temporal cadences.


\acknowledgments

The National Solar Observatory (NSO) is operated by the Association of Universities for Research in Astronomy, Inc. (AURA), under cooperative agreement with the National Science Foundation.  We extend thanks to Doug Gilliam at the Dunn Solar Telescope for help in instrument setup, alignment, and operation. Grants provided by the Major Research Instrument Program of the National Science Foundation, award number ATM04-21582 and ATM09-23560, supported the procurement of the IR camera and the high-efficiency bandpass isolation filters, which were used temporarily for the reported instrument. 


\bibliographystyle{spr-mp-sola-limited}
\bibliography{schad_manuscript}

\begin{thebibliography}{54}
\ifx\bisbn     \undefined \def\bisbn  #1{ISBN #1}\fi
\ifx\binits    \undefined \def\binits#1{#1}\fi
\ifx\bauthor   \undefined \def\bauthor#1{#1}\fi
\ifx\batitle   \undefined \def\batitle#1{#1}\fi
\ifx\bjtitle   \undefined \def\bjtitle#1{\textit{#1}}\fi
\ifx\bvolume   \undefined \def\bvolume#1{\textbf{#1}}\fi
\ifx\byear     \undefined \def\byear#1{#1}\fi
\ifx\bissue    \undefined \def\bissue#1{#1}\fi
\ifx\bfpage    \undefined \def\bfpage#1{#1}\fi
\ifx\blpage    \undefined \def\blpage #1{#1}\fi
\ifx\burl      \undefined \def\burl#1{\textsf{#1}}\fi
\ifx\href      \undefined \def\href#1#2{\textsf{#2}}\fi
\ifx\betal     \undefined \def\betal{\textit{et al.}}\fi
\ifx\bctitle   \undefined \def\bctitle#1{#1}\fi
\ifx\beditor   \undefined \def\beditor#1{#1}\fi
\ifx\bbtitle   \undefined \def\bbtitle#1{\textit{#1}}\fi
\ifx\bedition  \undefined \def\bedition#1{#1}\fi
\ifx\bseriesno \undefined \def\bseriesno#1{\textbf{#1}}\fi
\ifx\blocation \undefined \def\blocation#1{#1}\fi
\ifx\bsertitle \undefined \def\bsertitle#1{\textit{#1}}\fi
\ifx\bsnm      \undefined \def\bsnm#1{#1}\fi
\ifx\bsuffix   \undefined \def\bsuffix#1{#1}\fi
\ifx\bparticle \undefined \def\bparticle#1{#1}\fi
\ifx\barticle  \undefined \def\barticle#1{}\fi
\ifx\binstitute  \undefined \def\binstitute#1{#1}\fi
\ifx\bpublisher  \undefined \def\bpublisher#1{#1}\fi
\ifx\doiurl    \undefined
  \def\doiurl#1{\href{http://dx.doi.org/#1}{\textsf{DOI}}}\fi
\ifx\arxivurl  \undefined
  \def\arxivurl#1{\href{http://arxiv.org/abs/#1}{\textsf{arXiv}}}\fi
\ifx\adsurl    \undefined
  \def\adsurl#1{\href{http://adsabs.harvard.edu/abs/#1}{\textsf{ADS}}}\fi
\ifx\botherref \undefined \def\botherref#1{}\fi
\ifx\url       \undefined \def\url#1{\textsf{#1}}\fi
\ifx\bchapter  \undefined \def\bchapter#1{}\fi
\ifx\bbook     \undefined \def\bbook#1{}\fi
\ifx\bcomment  \undefined \def\bcomment#1{#1}\fi
\ifx\oauthor   \undefined \def\oauthor#1{#1}\fi
\ifx\citeauthoryear \undefined\def \citeauthoryear#1{#1}\fi
\def\endbibitem {}
\ifx\bconflocation  \undefined \def\bconflocation#1{#1} \fi

\bibitem[\protect\citeauthoryear{{Beckers}}{1998}]{beckers1998}
\begin{barticle}
\bauthor{\bsnm{{Beckers}}, \binits{J.M.}}:
\byear{1998},
\bjtitle{\aaps}
\bvolume{129},
\bfpage{191}.
\doiurl{10.1051/aas:1998180}.
\adsurl{1998A\%26AS..129..191B}.
\end{barticle}
\endbibitem

\bibitem[\protect\citeauthoryear{{Bershady}}{2009}]{bershady2009}
\begin{botherref}
\oauthor{\bsnm{{Bershady}}, \binits{M.A.}}:
2009,
\textit{ArXiv e-prints}.
\adsurl{2009arXiv0910.0167B}.
\end{botherref}
\endbibitem

\bibitem[\protect\citeauthoryear{{Breckinridge}}{1971}]{breckinridge1971}
\begin{barticle}
\bauthor{\bsnm{{Breckinridge}}, \binits{J.B.}}:
\byear{1971},
\bjtitle{\it Appl. Opt.}
\bvolume{10},
\bfpage{286}.
\doiurl{10.1364/AO.10.000286}.
\adsurl{1971ApOpt..10..286B}.
\end{barticle}
\endbibitem

\bibitem[\protect\citeauthoryear{{Calcines}
  \textit{et~al.}}{2014}]{calcines2014}
\begin{bchapter}
\bauthor{\bsnm{{Calcines}}, \binits{A.}},
\bauthor{\bsnm{{L{\'o}pez}}, \binits{R.L.}},
\bauthor{\bsnm{{Collados}}, \binits{M.}},
\bauthor{\bsnm{{Vega Reyes}}, \binits{N.}}:
\byear{2014},
In: \bbtitle{Ground-based and Airborne Instrumentation for Astronomy V},
\bsertitle{\procspie}
\bseriesno{9147},
\bfpage{91473I}.
\doiurl{10.1117/12.2053577}.
\adsurl{2014SPIE.9147E..3IC}.
\end{bchapter}
\endbibitem

\bibitem[\protect\citeauthoryear{{Cao} \textit{et~al.}}{2004}]{cao2004}
\begin{bchapter}
\bauthor{\bsnm{{Cao}}, \binits{W.}},
\bauthor{\bsnm{{Denker}}, \binits{C.J.}},
\bauthor{\bsnm{{Wang}}, \binits{H.}},
\bauthor{\bsnm{{Ma}}, \binits{J.}},
\bauthor{\bsnm{{Qu}}, \binits{M.}},
\bauthor{\bsnm{{Wang}}, \binits{J.}},
\bauthor{\bsnm{{Goode}}, \binits{P.R.}}:
\byear{2004},
In: \beditor{\bsnm{{Fineschi}}, \binits{S.}},
\beditor{\bsnm{{Gummin}}, \binits{M.A.}} (eds.)
\bbtitle{Telescopes and Instrumentation for Solar Astrophysics},
\bsertitle{\procspie}
\bseriesno{5171},
\bfpage{307}.
\doiurl{10.1117/12.506429}.
\adsurl{2004SPIE.5171..307C}.
\end{bchapter}
\endbibitem

\bibitem[\protect\citeauthoryear{{Cao} \textit{et~al.}}{2006}]{cao2006}
\begin{barticle}
\bauthor{\bsnm{{Cao}}, \binits{W.}},
\bauthor{\bsnm{{Jing}}, \binits{J.}},
\bauthor{\bsnm{{Ma}}, \binits{J.}},
\bauthor{\bsnm{{Xu}}, \binits{Y.}},
\bauthor{\bsnm{{Wang}}, \binits{H.}},
\bauthor{\bsnm{{Goode}}, \binits{P.R.}}:
\byear{2006},
\bjtitle{\pasp}
\bvolume{118},
\bfpage{838}.
\doiurl{10.1086/505408}.
\adsurl{2006PASP..118..838C}.
\end{barticle}
\endbibitem

\bibitem[\protect\citeauthoryear{{Cao} \textit{et~al.}}{2012}]{cao2012}
\begin{bchapter}
\bauthor{\bsnm{{Cao}}, \binits{W.}},
\bauthor{\bsnm{{Goode}}, \binits{P.R.}},
\bauthor{\bsnm{{Ahn}}, \binits{K.}},
\bauthor{\bsnm{{Gorceix}}, \binits{N.}},
\bauthor{\bsnm{{Schmidt}}, \binits{W.}},
\bauthor{\bsnm{{Lin}}, \binits{H.}}:
\byear{2012},
In: \beditor{\bsnm{{Rimmele}}, \binits{T.R.}},
\beditor{\bsnm{{Tritschler}}, \binits{A.}},
\beditor{\bsnm{{W{\"o}ger}}, \binits{F.}},
\beditor{\bsnm{{Collados Vera}}, \binits{M.}},
\beditor{\bsnm{{Socas-Navarro}}, \binits{H.}},
\beditor{\bsnm{{Schlichenmaier}}, \binits{R.}},
\beditor{\bsnm{{Carlsson}}, \binits{M.}},
\beditor{\bsnm{{Berger}}, \binits{T.}},
\beditor{\bsnm{{Cadavid}}, \binits{A.}},
\beditor{\bsnm{{Gilbert}}, \binits{P.R.}},
\beditor{\bsnm{{Goode}}, \binits{P.R.}},
\beditor{\bsnm{{Kn{\"o}lker}}, \binits{M.}} (eds.)
\bbtitle{Second ATST-EAST Meeting: Magnetic Fields from the Photosphere to the
  Corona.},
\bsertitle{Astronomical Society of the Pacific Conference Series}
\bseriesno{463},
\bfpage{291}.
\adsurl{2012ASPC..463..291C}.
\end{bchapter}
\endbibitem

\bibitem[\protect\citeauthoryear{{Casini}, {Judge}, and
  {Schad}}{2012}]{casini2012}
\begin{barticle}
\bauthor{\bsnm{{Casini}}, \binits{R.}},
\bauthor{\bsnm{{Judge}}, \binits{P.G.}},
\bauthor{\bsnm{{Schad}}, \binits{T.A.}}:
\byear{2012},
\bjtitle{\apj}
\bvolume{756},
\bfpage{194}.
\doiurl{10.1088/0004-637X/756/2/194}.
\adsurl{2012ApJ...756..194C}.
\end{barticle}
\endbibitem

\bibitem[\protect\citeauthoryear{{Cavallini}}{2006}]{cavallini2006}
\begin{barticle}
\bauthor{\bsnm{{Cavallini}}, \binits{F.}}:
\byear{2006},
\bjtitle{\solphys}
\bvolume{236},
\bfpage{415}.
\doiurl{10.1007/s11207-006-0103-8}.
\adsurl{2006SoPh..236..415C}.
\end{barticle}
\endbibitem

\bibitem[\protect\citeauthoryear{{Dekker} \textit{et~al.}}{2000}]{dekker2000}
\begin{bchapter}
\bauthor{\bsnm{{Dekker}}, \binits{H.}},
\bauthor{\bsnm{{D'Odorico}}, \binits{S.}},
\bauthor{\bsnm{{Kaufer}}, \binits{A.}},
\bauthor{\bsnm{{Delabre}}, \binits{B.}},
\bauthor{\bsnm{{Kotzlowski}}, \binits{H.}}:
\byear{2000},
In: \beditor{\bsnm{{Iye}}, \binits{M.}},
\beditor{\bsnm{{Moorwood}}, \binits{A.F.}} (eds.)
\bbtitle{Optical and IR Telescope Instrumentation and Detectors},
\bsertitle{\procspie}
\bseriesno{4008},
\bfpage{534}.
\doiurl{10.1117/12.395512}.
\adsurl{2000SPIE.4008..534D}.
\end{bchapter}
\endbibitem

\bibitem[\protect\citeauthoryear{{Doerr}, {von der L{\"u}he}, and
  {Kentischer}}{2008}]{doerr2008}
\begin{bchapter}
\bauthor{\bsnm{{Doerr}}, \binits{H.-P.}},
\bauthor{\bsnm{{von der L{\"u}he}}, \binits{O.} \bsuffix{II}},
\bauthor{\bsnm{{Kentischer}}, \binits{T.J.}}:
\byear{2008},
In: \bbtitle{Ground-based and Airborne Instrumentation for Astronomy II},
\bsertitle{\procspie}
\bseriesno{7014},
\bfpage{701417}.
\doiurl{10.1117/12.789180}.
\adsurl{2008SPIE.7014E..17D}.
\end{bchapter}
\endbibitem

\bibitem[\protect\citeauthoryear{{Dunn}}{1964}]{dunn1964}
\begin{barticle}
\bauthor{\bsnm{{Dunn}}, \binits{R.B.}}:
\byear{1964},
\bjtitle{\it Appl. Opt.}
\bvolume{3},
\bfpage{1353}.
\doiurl{10.1364/AO.3.001353}.
\adsurl{1964ApOpt...3.1353D}.
\end{barticle}
\endbibitem

\bibitem[\protect\citeauthoryear{{Dunn} and {Smartt}}{1991}]{dunn1991}
\begin{barticle}
\bauthor{\bsnm{{Dunn}}, \binits{R.B.}},
\bauthor{\bsnm{{Smartt}}, \binits{R.N.}}:
\byear{1991},
\bjtitle{Advances in Space Research}
\bvolume{11},
\bfpage{139}.
\doiurl{10.1016/0273-1177(91)90371-P}.
\adsurl{1991AdSpR..11..139D}.
\end{barticle}
\endbibitem

\bibitem[\protect\citeauthoryear{{Freeland} and {Handy}}{1998}]{freeland1998}
\begin{barticle}
\bauthor{\bsnm{{Freeland}}, \binits{S.L.}},
\bauthor{\bsnm{{Handy}}, \binits{B.N.}}:
\byear{1998},
\bjtitle{\solphys}
\bvolume{182},
\bfpage{497}.
\doiurl{10.1023/A:1005038224881}.
\adsurl{1998SoPh..182..497F}.
\end{barticle}
\endbibitem

\bibitem[\protect\citeauthoryear{{Harvey} and {Hall}}{1971}]{harvey1971}
\begin{bchapter}
\bauthor{\bsnm{{Harvey}}, \binits{J.}},
\bauthor{\bsnm{{Hall}}, \binits{D.}}:
\byear{1971},
In: \beditor{\bsnm{{Howard}}, \binits{R.}} (ed.)
\bbtitle{Solar Magnetic Fields},
\bsertitle{IAU Symposium}
\bseriesno{43},
\bfpage{279}.
\adsurl{1971IAUS...43..279H}.
\end{bchapter}
\endbibitem

\bibitem[\protect\citeauthoryear{{Jaeggli} \textit{et~al.}}{2010}]{jaeggli2010}
\begin{barticle}
\bauthor{\bsnm{{Jaeggli}}, \binits{S.A.}},
\bauthor{\bsnm{{Lin}}, \binits{H.}},
\bauthor{\bsnm{{Mickey}}, \binits{D.L.}},
\bauthor{\bsnm{{Kuhn}}, \binits{J.R.}},
\bauthor{\bsnm{{Hegwer}}, \binits{S.L.}},
\bauthor{\bsnm{{Rimmele}}, \binits{T.R.}},
\bauthor{\bsnm{{Penn}}, \binits{M.J.}}:
\byear{2010},
\bjtitle{\memsai}
\bvolume{81},
\bfpage{763}.
\adsurl{2010MmSAI..81..763J}.
\end{barticle}
\endbibitem

\bibitem[\protect\citeauthoryear{{Johanneson}
  \textit{et~al.}}{1992}]{johanneson1992}
\begin{barticle}
\bauthor{\bsnm{{Johanneson}}, \binits{A.}},
\bauthor{\bsnm{{Bida}}, \binits{T.}},
\bauthor{\bsnm{{Lites}}, \binits{B.}},
\bauthor{\bsnm{{Scharmer}}, \binits{G.B.}}:
\byear{1992},
\bjtitle{\aap}
\bvolume{258},
\bfpage{572}.
\adsurl{1992A\%26A...258..572J}.
\end{barticle}
\endbibitem

\bibitem[\protect\citeauthoryear{{Katsukawa}
  \textit{et~al.}}{2016}]{katsukawa2016}
\begin{bchapter}
\bauthor{\bsnm{{Katsukawa}}, \binits{Y.}},
\bauthor{\bsnm{{Kamata}}, \binits{Y.}},
\bauthor{\bsnm{{Anan}}, \binits{T.}},
\bauthor{\bsnm{{Hara}}, \binits{H.}},
\bauthor{\bsnm{{Suematsu}}, \binits{Y.}},
\bauthor{\bsnm{{Bando}}, \binits{T.}},
\bauthor{\bsnm{{Ichimoto}}, \binits{K.}},
\bauthor{\bsnm{{Shimizu}}, \binits{T.}}:
\byear{2016},
In: \bbtitle{Space Telescopes and Instrumentation 2016: Optical, Infrared, and
  Millimeter Wave},
\bsertitle{\procspie}
\bseriesno{9904},
\bfpage{99045I}.
\doiurl{10.1117/12.2234547}.
\adsurl{2016SPIE.9904E..5IK}.
\end{bchapter}
\endbibitem

\bibitem[\protect\citeauthoryear{{Keller} and {Johannesson}}{1995}]{keller1995}
\begin{barticle}
\bauthor{\bsnm{{Keller}}, \binits{C.U.}},
\bauthor{\bsnm{{Johannesson}}, \binits{A.}}:
\byear{1995},
\bjtitle{\aaps}
\bvolume{110},
\bfpage{565}.
\adsurl{1995A\%26AS..110..565K}.
\end{barticle}
\endbibitem

\bibitem[\protect\citeauthoryear{Kentischer
  \textit{et~al.}}{2012}]{kentischer2012}
\begin{bchapter}
\bauthor{\bsnm{Kentischer}, \binits{T.J.}},
\bauthor{\bsnm{Schmidt}, \binits{W..}},
\bauthor{\bparticle{von~der} \bsnm{L{\"u}he}, \binits{O.}},
\bauthor{\bsnm{Sigwarth}, \binits{M.}},
\bauthor{\bsnm{Bell}, \binits{A.}},
\bauthor{\bsnm{Halbgewachs}, \binits{C.}},
\bauthor{\bsnm{Fischer}, \binits{A.}}:
\byear{2012},
\bctitle{The visible tunable filtergraph for the atst}.
\bseriesno{8446},
\bfpage{844677}.
\doiurl{10.1117/12.926789}.
\burl{http://dx.doi.org/10.1117/12.926789}.
\end{bchapter}
\endbibitem

\bibitem[\protect\citeauthoryear{{Kleint} and {Gandorfer}}{2015}]{kleint2015}
\begin{botherref}
\oauthor{\bsnm{{Kleint}}, \binits{L.}},
\oauthor{\bsnm{{Gandorfer}}, \binits{A.}}:
2015,
\textit{\ssr}.
\doiurl{10.1007/s11214-015-0208-1}.
\adsurl{2015SSRv..tmp..114K}.
\end{botherref}
\endbibitem

\bibitem[\protect\citeauthoryear{{Kuckein}, {Mart{\'{\i}}nez Pillet}, and
  {Centeno}}{2012}]{kuckein2012}
\begin{barticle}
\bauthor{\bsnm{{Kuckein}}, \binits{C.}},
\bauthor{\bsnm{{Mart{\'{\i}}nez Pillet}}, \binits{V.}},
\bauthor{\bsnm{{Centeno}}, \binits{R.}}:
\byear{2012},
\bjtitle{\aap}
\bvolume{542},
\bfpage{A112}.
\doiurl{10.1051/0004-6361/201218887}.
\adsurl{2012A\%26A...542A.112K}.
\end{barticle}
\endbibitem

\bibitem[\protect\citeauthoryear{{Lagg} \textit{et~al.}}{2015}]{lagg2015}
\begin{botherref}
\oauthor{\bsnm{{Lagg}}, \binits{A.}},
\oauthor{\bsnm{{Lites}}, \binits{B.}},
\oauthor{\bsnm{{Harvey}}, \binits{J.}},
\oauthor{\bsnm{{Gosain}}, \binits{S.}},
\oauthor{\bsnm{{Centeno}}, \binits{R.}}:
2015,
\textit{\ssr}.
\doiurl{10.1007/s11214-015-0219-y}.
\adsurl{2015SSRv..tmp..115L}.
\end{botherref}
\endbibitem

\bibitem[\protect\citeauthoryear{Lee and Schachter}{1980}]{lee1980}
\begin{barticle}
\bauthor{\bsnm{Lee}, \binits{D.T.}},
\bauthor{\bsnm{Schachter}, \binits{B.J.}}:
\byear{1980},
\bjtitle{International Journal of Computer {\&} Information Sciences}
\bvolume{9}(\bissue{3}),
\bfpage{219}.
\doiurl{10.1007/BF00977785}.
\burl{http://dx.doi.org/10.1007/BF00977785}.
\end{barticle}
\endbibitem

\bibitem[\protect\citeauthoryear{{Leenaarts}
  \textit{et~al.}}{2016}]{leenaarts2016}
\begin{barticle}
\bauthor{\bsnm{{Leenaarts}}, \binits{J.}},
\bauthor{\bsnm{{Golding}}, \binits{T.}},
\bauthor{\bsnm{{Carlsson}}, \binits{M.}},
\bauthor{\bsnm{{Libbrecht}}, \binits{T.}},
\bauthor{\bsnm{{Joshi}}, \binits{J.}}:
\byear{2016},
\bjtitle{\aap}
\bvolume{594},
\bfpage{A104}.
\doiurl{10.1051/0004-6361/201628490}.
\adsurl{2016A\%26A...594A.104L}.
\end{barticle}
\endbibitem

\bibitem[\protect\citeauthoryear{{Lemen} \textit{et~al.}}{2011}]{lemen2011}
\begin{botherref}
\oauthor{\bsnm{{Lemen}}, \binits{J.R.}},
\oauthor{\bsnm{{Title}}, \binits{A.M.}},
\oauthor{\bsnm{{Akin}}, \binits{D.J.}},
\oauthor{\bsnm{{Boerner}}, \binits{P.F.}},
\oauthor{\bsnm{{Chou}}, \binits{C.}},
\oauthor{\bsnm{{Drake}}, \binits{J.F.}},
\oauthor{\bsnm{{Duncan}}, \binits{D.W.}},
\oauthor{\bsnm{{Edwards}}, \binits{C.G.}},
\oauthor{\bsnm{{Friedlaender}}, \binits{F.M.}},
\oauthor{\bsnm{{Heyman}}, \binits{G.F.}},
\oauthor{\bsnm{{Hurlburt}}, \binits{N.E.}},
\oauthor{\bsnm{{Katz}}, \binits{N.L.}},
\oauthor{\bsnm{{Kushner}}, \binits{G.D.}},
\oauthor{\bsnm{{Levay}}, \binits{M.}},
\oauthor{\bsnm{{Lindgren}}, \binits{R.W.}},
\oauthor{\bsnm{{Mathur}}, \binits{D.P.}},
\oauthor{\bsnm{{McFeaters}}, \binits{E.L.}},
\oauthor{\bsnm{{Mitchell}}, \binits{S.}},
\oauthor{\bsnm{{Rehse}}, \binits{R.A.}},
\oauthor{\bsnm{{Schrijver}}, \binits{C.J.}},
\oauthor{\bsnm{{Springer}}, \binits{L.A.}},
\oauthor{\bsnm{{Stern}}, \binits{R.A.}},
\oauthor{\bsnm{{Tarbell}}, \binits{T.D.}},
\oauthor{\bsnm{{Wuelser}}, \binits{J.-P.}},
\oauthor{\bsnm{{Wolfson}}, \binits{C.J.}},
\oauthor{\bsnm{{Yanari}}, \binits{C.}},
\oauthor{\bsnm{{Bookbinder}}, \binits{J.A.}},
\oauthor{\bsnm{{Cheimets}}, \binits{P.N.}},
\oauthor{\bsnm{{Caldwell}}, \binits{D.}},
\oauthor{\bsnm{{Deluca}}, \binits{E.E.}},
\oauthor{\bsnm{{Gates}}, \binits{R.}},
\oauthor{\bsnm{{Golub}}, \binits{L.}},
\oauthor{\bsnm{{Park}}, \binits{S.}},
\oauthor{\bsnm{{Podgorski}}, \binits{W.A.}},
\oauthor{\bsnm{{Bush}}, \binits{R.I.}},
\oauthor{\bsnm{{Scherrer}}, \binits{P.H.}},
\oauthor{\bsnm{{Gummin}}, \binits{M.A.}},
\oauthor{\bsnm{{Smith}}, \binits{P.}},
\oauthor{\bsnm{{Auker}}, \binits{G.}},
\oauthor{\bsnm{{Jerram}}, \binits{P.}},
\oauthor{\bsnm{{Pool}}, \binits{P.}},
\oauthor{\bsnm{{Soufli}}, \binits{R.}},
\oauthor{\bsnm{{Windt}}, \binits{D.L.}},
\oauthor{\bsnm{{Beardsley}}, \binits{S.}},
\oauthor{\bsnm{{Clapp}}, \binits{M.}},
\oauthor{\bsnm{{Lang}}, \binits{J.}},
\oauthor{\bsnm{{Waltham}}, \binits{N.}}:
2011,
\textit{\solphys},
172.
\doiurl{10.1007/s11207-011-9776-8}.
\adsurl{2011SoPh..tmp..172L}.
\end{botherref}
\endbibitem

\bibitem[\protect\citeauthoryear{Lin}{2014}]{lin2014}
\begin{bchapter}
\bauthor{\bsnm{Lin}, \binits{H.}}:
\byear{2014},
\bctitle{mxspec: a massively multiplexed full-disk spectroheliograph for solar
  physics research}.
\bseriesno{9147},
\bfpage{914712}.
\doiurl{10.1117/12.2057120}.
\burl{http://dx.doi.org/10.1117/12.2057120}.
\end{bchapter}
\endbibitem

\bibitem[\protect\citeauthoryear{{Lin}}{2016}]{lin2016}
\begin{barticle}
\bauthor{\bsnm{{Lin}}, \binits{H.}}:
\byear{2016},
\bjtitle{Frontiers in Astronomy and Space Sciences}
\bvolume{3},
\bfpage{9}.
\doiurl{10.3389/fspas.2016.00009}.
\adsurl{2016FrASS...3....9L}.
\end{barticle}
\endbibitem

\bibitem[\protect\citeauthoryear{{Lin} and {Versteegh}}{2006}]{lin2006}
\begin{bchapter}
\bauthor{\bsnm{{Lin}}, \binits{H.}},
\bauthor{\bsnm{{Versteegh}}, \binits{A.}}:
\byear{2006},
In: \bbtitle{Society of Photo-Optical Instrumentation Engineers (SPIE)
  Conference Series},
\bsertitle{\procspie}
\bseriesno{6269},
\bfpage{62690K}.
\doiurl{10.1117/12.670852}.
\adsurl{2006SPIE.6269E..0KL}.
\end{bchapter}
\endbibitem

\bibitem[\protect\citeauthoryear{{Martin} \textit{et~al.}}{1974}]{martin1974}
\begin{barticle}
\bauthor{\bsnm{{Martin}}, \binits{S.F.}},
\bauthor{\bsnm{{Ramsey}}, \binits{H.E.}},
\bauthor{\bsnm{{Carroll}}, \binits{G.A.}},
\bauthor{\bsnm{{Martin}}, \binits{D.C.}}:
\byear{1974},
\bjtitle{\solphys}
\bvolume{37},
\bfpage{343}.
\doiurl{10.1007/BF00152493}.
\adsurl{1974SoPh...37..343M}.
\end{barticle}
\endbibitem

\bibitem[\protect\citeauthoryear{{November} and {Simon}}{1988}]{november1988}
\begin{barticle}
\bauthor{\bsnm{{November}}, \binits{L.J.}},
\bauthor{\bsnm{{Simon}}, \binits{G.W.}}:
\byear{1988},
\bjtitle{\apj}
\bvolume{333},
\bfpage{427}.
\doiurl{10.1086/166758}.
\adsurl{1988ApJ...333..427N}.
\end{barticle}
\endbibitem

\bibitem[\protect\citeauthoryear{Penn}{2014}]{penn2014}
\begin{barticle}
\bauthor{\bsnm{Penn}, \binits{M.J.}}:
\byear{2014},
\bjtitle{Living Reviews in Solar Physics}
\bvolume{11}(\bissue{1}),
\bfpage{2}.
\doiurl{10.12942/lrsp-2014-2}.
\burl{http://dx.doi.org/10.12942/lrsp-2014-2}.
\end{barticle}
\endbibitem

\bibitem[\protect\citeauthoryear{{Pilachowski}
  \textit{et~al.}}{1995}]{pilachowski1995}
\begin{barticle}
\bauthor{\bsnm{{Pilachowski}}, \binits{C.}},
\bauthor{\bsnm{{Dekker}}, \binits{H.}},
\bauthor{\bsnm{{Hinkle}}, \binits{K.}},
\bauthor{\bsnm{{Tull}}, \binits{R.}},
\bauthor{\bsnm{{Vogt}}, \binits{S.}},
\bauthor{\bsnm{{Walker}}, \binits{D.D.}},
\bauthor{\bsnm{{Diego}}, \binits{F.}},
\bauthor{\bsnm{{Angel}}, \binits{R.}}:
\byear{1995},
\bjtitle{\pasp}
\bvolume{107},
\bfpage{983}.
\doiurl{10.1086/133648}.
\adsurl{1995PASP..107..983P}.
\end{barticle}
\endbibitem

\bibitem[\protect\citeauthoryear{{Reardon} and {Cavallini}}{2008}]{reardon2008}
\begin{barticle}
\bauthor{\bsnm{{Reardon}}, \binits{K.P.}},
\bauthor{\bsnm{{Cavallini}}, \binits{F.}}:
\byear{2008},
\bjtitle{\aap}
\bvolume{481},
\bfpage{897}.
\doiurl{10.1051/0004-6361:20078473}.
\adsurl{2008A\%26A...481..897R}.
\end{barticle}
\endbibitem

\bibitem[\protect\citeauthoryear{{Rimmele}}{1994}]{rimmele1994}
\begin{barticle}
\bauthor{\bsnm{{Rimmele}}, \binits{T.R.}}:
\byear{1994},
\bjtitle{\aap}
\bvolume{290},
\bfpage{972}.
\adsurl{1994A\%26A...290..972R}.
\end{barticle}
\endbibitem

\bibitem[\protect\citeauthoryear{Rimmele and Marino}{2011}]{rimmele2011}
\begin{barticle}
\bauthor{\bsnm{Rimmele}, \binits{T.R.}},
\bauthor{\bsnm{Marino}, \binits{J.}}:
\byear{2011},
\bjtitle{Living Reviews in Solar Physics}
\bvolume{8}(\bissue{1}),
\bfpage{2}.
\doiurl{10.12942/lrsp-2011-2}.
\burl{http://dx.doi.org/10.12942/lrsp-2011-2}.
\end{barticle}
\endbibitem

\bibitem[\protect\citeauthoryear{{Rimmele} \textit{et~al.}}{2004}]{rimmele2004}
\begin{bchapter}
\bauthor{\bsnm{{Rimmele}}, \binits{T.R.}},
\bauthor{\bsnm{{Richards}}, \binits{K.}},
\bauthor{\bsnm{{Hegwer}}, \binits{S.}},
\bauthor{\bsnm{{Fletcher}}, \binits{S.}},
\bauthor{\bsnm{{Gregory}}, \binits{S.}},
\bauthor{\bsnm{{Moretto}}, \binits{G.}},
\bauthor{\bsnm{{Didkovsky}}, \binits{L.V.}},
\bauthor{\bsnm{{Denker}}, \binits{C.J.}},
\bauthor{\bsnm{{Dolgushin}}, \binits{A.}},
\bauthor{\bsnm{{Goode}}, \binits{P.R.}},
\bauthor{\bsnm{{Langlois}}, \binits{M.}},
\bauthor{\bsnm{{Marino}}, \binits{J.}},
\bauthor{\bsnm{{Marquette}}, \binits{W.}}:
\byear{2004},
In: \beditor{\bsnm{{Fineschi}}, \binits{S.}},
\beditor{\bsnm{{Gummin}}, \binits{M.A.}} (eds.)
\bbtitle{Society of Photo-Optical Instrumentation Engineers (SPIE) Conference
  Series},
\bsertitle{Society of Photo-Optical Instrumentation Engineers (SPIE) Conference
  Series}
\bseriesno{5171},
\bfpage{179}.
\doiurl{10.1117/12.508513}.
\adsurl{2004SPIE.5171..179R}.
\end{bchapter}
\endbibitem

\bibitem[\protect\citeauthoryear{{Rimmele} \textit{et~al.}}{2015}]{rimmele2015}
\begin{barticle}
\bauthor{\bsnm{{Rimmele}}, \binits{T.}},
\bauthor{\bsnm{{McMullin}}, \binits{J.}},
\bauthor{\bsnm{{Warner}}, \binits{M.}},
\bauthor{\bsnm{{Craig}}, \binits{S.}},
\bauthor{\bsnm{{Woeger}}, \binits{F.}},
\bauthor{\bsnm{{Tritschler}}, \binits{A.}},
\bauthor{\bsnm{{Cassini}}, \binits{R.}},
\bauthor{\bsnm{{Kuhn}}, \binits{J.}},
\bauthor{\bsnm{{Lin}}, \binits{H.}},
\bauthor{\bsnm{{Schmidt}}, \binits{W.}},
\bauthor{\bsnm{{Berukoff}}, \binits{S.}},
\bauthor{\bsnm{{Reardon}}, \binits{K.}},
\bauthor{\bsnm{{Goode}}, \binits{P.}},
\bauthor{\bsnm{{Knoelker}}, \binits{M.}},
\bauthor{\bsnm{{Rosner}}, \binits{R.}},
\bauthor{\bsnm{{Mathioudakis}}, \binits{M.}},
\bauthor{\bsnm{{DKIST TEAM}}}:
\byear{2015},
\bjtitle{IAU General Assembly}
\bvolume{22},
\bfpage{2255176}.
\adsurl{2015IAUGA..2255176R}.
\end{barticle}
\endbibitem

\bibitem[\protect\citeauthoryear{{Schad}}{2016}]{schad2016agu}
\begin{botherref}
\oauthor{\bsnm{{Schad}}, \binits{T.A.}}:
2016,
\textit{AGU Fall Meeting Abstracts}.
\end{botherref}
\endbibitem

\bibitem[\protect\citeauthoryear{{Schad} \textit{et~al.}}{2016}]{schad2016}
\begin{barticle}
\bauthor{\bsnm{{Schad}}, \binits{T.A.}},
\bauthor{\bsnm{{Penn}}, \binits{M.J.}},
\bauthor{\bsnm{{Lin}}, \binits{H.}},
\bauthor{\bsnm{{Judge}}, \binits{P.G.}}:
\byear{2016},
\bjtitle{\apj}
\bvolume{833},
\bfpage{5}.
\doiurl{10.3847/0004-637X/833/1/5}.
\adsurl{2016ApJ...833....5S}.
\end{barticle}
\endbibitem

\bibitem[\protect\citeauthoryear{{Schad} \textit{et~al.}}{2014}]{schad2014}
\begin{bchapter}
\bauthor{\bsnm{{Schad}}, \binits{T.}},
\bauthor{\bsnm{{Lin}}, \binits{H.}},
\bauthor{\bsnm{{Ichimoto}}, \binits{K.}},
\bauthor{\bsnm{{Katsukawa}}, \binits{Y.}}:
\byear{2014},
In: \bbtitle{Ground-based and Airborne Instrumentation for Astronomy V},
\bsertitle{\procspie}
\bseriesno{9147},
\bfpage{91476E}.
\doiurl{10.1117/12.2057125}.
\adsurl{2014SPIE.9147E..6ES}.
\end{bchapter}
\endbibitem

\bibitem[\protect\citeauthoryear{{Scharmer}
  \textit{et~al.}}{2008}]{scharmer2008}
\begin{barticle}
\bauthor{\bsnm{{Scharmer}}, \binits{G.B.}},
\bauthor{\bsnm{{Narayan}}, \binits{G.}},
\bauthor{\bsnm{{Hillberg}}, \binits{T.}},
\bauthor{\bsnm{{de la Cruz Rodriguez}}, \binits{J.}},
\bauthor{\bsnm{{L{\"o}fdahl}}, \binits{M.G.}},
\bauthor{\bsnm{{Kiselman}}, \binits{D.}},
\bauthor{\bsnm{{S{\"u}tterlin}}, \binits{P.}},
\bauthor{\bsnm{{van Noort}}, \binits{M.}},
\bauthor{\bsnm{{Lagg}}, \binits{A.}}:
\byear{2008},
\bjtitle{\apjl}
\bvolume{689},
\bfpage{L69}.
\doiurl{10.1086/595744}.
\adsurl{2008ApJ...689L..69S}.
\end{barticle}
\endbibitem

\bibitem[\protect\citeauthoryear{{Scherrer}
  \textit{et~al.}}{2012}]{scherrer2012}
\begin{barticle}
\bauthor{\bsnm{{Scherrer}}, \binits{P.H.}},
\bauthor{\bsnm{{Schou}}, \binits{J.}},
\bauthor{\bsnm{{Bush}}, \binits{R.I.}},
\bauthor{\bsnm{{Kosovichev}}, \binits{A.G.}},
\bauthor{\bsnm{{Bogart}}, \binits{R.S.}},
\bauthor{\bsnm{{Hoeksema}}, \binits{J.T.}},
\bauthor{\bsnm{{Liu}}, \binits{Y.}},
\bauthor{\bsnm{{Duvall}}, \binits{T.L.}},
\bauthor{\bsnm{{Zhao}}, \binits{J.}},
\bauthor{\bsnm{{Title}}, \binits{A.M.}},
\bauthor{\bsnm{{Schrijver}}, \binits{C.J.}},
\bauthor{\bsnm{{Tarbell}}, \binits{T.D.}},
\bauthor{\bsnm{{Tomczyk}}, \binits{S.}}:
\byear{2012},
\bjtitle{\solphys}
\bvolume{275},
\bfpage{207}.
\doiurl{10.1007/s11207-011-9834-2}.
\adsurl{2012SoPh..275..207S}.
\end{barticle}
\endbibitem

\bibitem[\protect\citeauthoryear{{Schmidt} \textit{et~al.}}{2017}]{schmidt2017}
\begin{barticle}
\bauthor{\bsnm{{Schmidt}}, \binits{D.}},
\bauthor{\bsnm{{Gorceix}}, \binits{N.}},
\bauthor{\bsnm{{Goode}}, \binits{P.R.}},
\bauthor{\bsnm{{Marino}}, \binits{J.}},
\bauthor{\bsnm{{Rimmele}}, \binits{T.}},
\bauthor{\bsnm{{Berkefeld}}, \binits{T.}},
\bauthor{\bsnm{{W{\"o}ger}}, \binits{F.}},
\bauthor{\bsnm{{Zhang}}, \binits{X.}},
\bauthor{\bsnm{{Rigaut}}, \binits{F.}},
\bauthor{\bsnm{{von der L{\"u}he}}, \binits{O.}}:
\byear{2017},
\bjtitle{\aap}
\bvolume{597},
\bfpage{L8}.
\doiurl{10.1051/0004-6361/201629970}.
\adsurl{2017A\%26A...597L...8S}.
\end{barticle}
\endbibitem

\bibitem[\protect\citeauthoryear{{Schmidt} \textit{et~al.}}{2016}]{schmidt2016}
\begin{bchapter}
\bauthor{\bsnm{{Schmidt}}, \binits{W.}},
\bauthor{\bsnm{{Schubert}}, \binits{M.}},
\bauthor{\bsnm{{Ellwarth}}, \binits{M.}},
\bauthor{\bsnm{{Baumgartner}}, \binits{J.}},
\bauthor{\bsnm{{Bell}}, \binits{A.}},
\bauthor{\bsnm{{Fischer}}, \binits{A.}},
\bauthor{\bsnm{{Halbgewachs}}, \binits{C.}},
\bauthor{\bsnm{{Heidecke}}, \binits{F.}},
\bauthor{\bsnm{{Kentischer}}, \binits{T.}},
\bauthor{\bsnm{{von der L{\"u}he}}, \binits{O.}},
\bauthor{\bsnm{{Scheiffelen}}, \binits{T.}},
\bauthor{\bsnm{{Sigwarth}}, \binits{M.}}:
\byear{2016},
In: \bbtitle{Society of Photo-Optical Instrumentation Engineers (SPIE)
  Conference Series},
\bsertitle{\procspie}
\bseriesno{9908},
\bfpage{99084N}.
\doiurl{10.1117/12.2232518}.
\adsurl{2016SPIE.9908E..4NS}.
\end{bchapter}
\endbibitem

\bibitem[\protect\citeauthoryear{{Semel}}{2003}]{semel2003}
\begin{barticle}
\bauthor{\bsnm{{Semel}}, \binits{M.}}:
\byear{2003},
\bjtitle{\aap}
\bvolume{401},
\bfpage{1}.
\doiurl{10.1051/0004-6361:20021606}.
\adsurl{2003A\%26A...401....1S}.
\end{barticle}
\endbibitem

\bibitem[\protect\citeauthoryear{Sigwarth \textit{et~al.}}{2016}]{sigwarth2016}
\begin{bchapter}
\bauthor{\bsnm{Sigwarth}, \binits{M.}},
\bauthor{\bsnm{Baumgartner}, \binits{J.}},
\bauthor{\bsnm{Bell}, \binits{A.}},
\bauthor{\bsnm{Cagnoli}, \binits{G.}},
\bauthor{\bsnm{Fischer}, \binits{A.}},
\bauthor{\bsnm{Halbgewachs}, \binits{C.}},
\bauthor{\bsnm{Heidecke}, \binits{F.}},
\bauthor{\bsnm{Kentischer}, \binits{T.J.}},
\bauthor{\bsnm{Kestner}, \binits{B.}},
\bauthor{\bsnm{Kuschnir}, \binits{P.}},
\bauthor{\bparticle{von~der} \bsnm{Lühe}, \binits{O.}},
\bauthor{\bsnm{Pinard}, \binits{L.}},
\bauthor{\bsnm{Michel}, \binits{C.}},
\bauthor{\bsnm{Reichman}, \binits{W.J.}},
\bauthor{\bsnm{Sassolas}, \binits{B.}},
\bauthor{\bsnm{Scheiffelen}, \binits{T.}},
\bauthor{\bsnm{Schmidt}, \binits{W.}}:
\byear{2016},
\bctitle{Development of high reflectivity coatings for large format fabry-perot
  etalons}.
\bseriesno{9908},
\bfpage{99084F}.
\doiurl{10.1117/12.2232271}.
\burl{http://dx.doi.org/10.1117/12.2232271}.
\end{bchapter}
\endbibitem

\bibitem[\protect\citeauthoryear{{Socas-Navarro}}{2010}]{socas_navarro2010}
\begin{barticle}
\bauthor{\bsnm{{Socas-Navarro}}, \binits{H.}}:
\byear{2010},
\bjtitle{Astronomische Nachrichten}
\bvolume{331},
\bfpage{581}.
\doiurl{10.1002/asna.201011377}.
\adsurl{2010AN....331..581S}.
\end{barticle}
\endbibitem

\bibitem[\protect\citeauthoryear{{Srivastava} and
  {Mathew}}{1999}]{srivastava1999}
\begin{barticle}
\bauthor{\bsnm{{Srivastava}}, \binits{N.}},
\bauthor{\bsnm{{Mathew}}, \binits{S.K.}}:
\byear{1999},
\bjtitle{\solphys}
\bvolume{185},
\bfpage{61}.
\doiurl{10.1023/A:1005189319845}.
\adsurl{1999SoPh..185...61S}.
\end{barticle}
\endbibitem

\bibitem[\protect\citeauthoryear{{Suematsu}
  \textit{et~al.}}{2016}]{suematsu2016}
\begin{bchapter}
\bauthor{\bsnm{{Suematsu}}, \binits{Y.}},
\bauthor{\bsnm{{Saito}}, \binits{K.}},
\bauthor{\bsnm{{Koyama}}, \binits{M.}},
\bauthor{\bsnm{{Enokida}}, \binits{Y.}},
\bauthor{\bsnm{{Okura}}, \binits{Y.}},
\bauthor{\bsnm{{Nakayasu}}, \binits{T.}},
\bauthor{\bsnm{{Sukegawa}}, \binits{T.}}:
\byear{2016},
In: \bbtitle{Space Telescopes and Instrumentation 2016: Optical, Infrared, and
  Millimeter Wave},
\bsertitle{\procspie}
\bseriesno{9904},
\bfpage{990411}.
\doiurl{10.1117/12.2231947}.
\adsurl{2016SPIE.9904E..11S}.
\end{bchapter}
\endbibitem

\bibitem[\protect\citeauthoryear{{van Noort}, {Rouppe van der Voort}, and
  {L{\"o}fdahl}}{2005}]{van_noort2005}
\begin{barticle}
\bauthor{\bsnm{{van Noort}}, \binits{M.}},
\bauthor{\bsnm{{Rouppe van der Voort}}, \binits{L.}},
\bauthor{\bsnm{{L{\"o}fdahl}}, \binits{M.G.}}:
\byear{2005},
\bjtitle{\solphys}
\bvolume{228},
\bfpage{191}.
\doiurl{10.1007/s11207-005-5782-z}.
\adsurl{2005SoPh..228..191V}.
\end{barticle}
\endbibitem

\bibitem[\protect\citeauthoryear{{Vissers} and {Rouppe van der
  Voort}}{2012}]{vissers2012}
\begin{barticle}
\bauthor{\bsnm{{Vissers}}, \binits{G.}},
\bauthor{\bsnm{{Rouppe van der Voort}}, \binits{L.}}:
\byear{2012},
\bjtitle{\apj}
\bvolume{750},
\bfpage{22}.
\doiurl{10.1088/0004-637X/750/1/22}.
\adsurl{2012ApJ...750...22V}.
\end{barticle}
\endbibitem

\bibitem[\protect\citeauthoryear{{von der L{\"u}he} and
  {Kentischer}}{2000}]{von_der_luhe2000}
\begin{barticle}
\bauthor{\bsnm{{von der L{\"u}he}}, \binits{O.}},
\bauthor{\bsnm{{Kentischer}}, \binits{T.J.}}:
\byear{2000},
\bjtitle{\aaps}
\bvolume{146},
\bfpage{499}.
\doiurl{10.1051/aas:2000283}.
\adsurl{2000A\%26AS..146..499V}.
\end{barticle}
\endbibitem

\bibitem[\protect\citeauthoryear{{W{\"o}ger}, {von der L{\"u}he}, and
  {Reardon}}{2008}]{woeger2008}
\begin{barticle}
\bauthor{\bsnm{{W{\"o}ger}}, \binits{F.}},
\bauthor{\bsnm{{von der L{\"u}he}}, \binits{O.}},
\bauthor{\bsnm{{Reardon}}, \binits{K.}}:
\byear{2008},
\bjtitle{\aap}
\bvolume{488},
\bfpage{375}.
\doiurl{10.1051/0004-6361:200809894}.
\adsurl{2008A\%26A...488..375W}.
\end{barticle}
\endbibitem

\end{thebibliography}


\end{article}

\end{document}